\documentclass[fleqn,usenatbib]{mnras}

\usepackage{newtxtext,newtxmath}
\usepackage{orcidlink}
\usepackage[T1]{fontenc}
\usepackage{booktabs}
\DeclareRobustCommand{\VAN}[3]{#2}
\let\VANthebibliography\thebibliography
\def\thebibliography{\DeclareRobustCommand{\VAN}[3]{##3}\VANthebibliography}

\usepackage{graphicx}	
\usepackage{amsmath}	
\usepackage{textcomp}
\usepackage{xcolor}
\graphicspath{{./}{image/}}
\usepackage{amsmath}
\def \firstRevise {}

\def \secRevise {}

\def \teamRevise {}
\def \referee {}

\title[MAGPI: evolution and drivers of gas turbulence]{The MAGPI Survey: the evolution and drivers of gas turbulence in intermediate-redshift galaxies}

\author[Y. Mai et al.]{
Yifan Mai$^{\orcidlink{0000-0003-3514-6280}}$,$^{1,2}$\thanks{E-mail: yifan.mai@sydney.edu.au}
Scott M. Croom$^{\orcidlink{0000-0003-2880-9197}}$,$^{1,2}$
Emily Wisnioski$^{\orcidlink{0000-0003-1657-7878}}$,$^{2,3}$
Sam P. Vaughan$^{\orcidlink{0000-0003-2265-7727}}$,$^{2,4,5,6}$
Mathew R. Varidel$^{\orcidlink{0000-0002-1648-8317}}$,$^{7}$
\newauthor
Andrew J. Battisti$^{\orcidlink{0000-0003-4569-2285}}$,$^{2,3}$
J. Trevor Mendel$^{\orcidlink{0000-0002-6327-9147}}$,$^{2,3}$
Marcie Mun$^{\orcidlink{0000-0002-3706-9955}}$,$^{2,3}$
Takafumi Tsukui$^{\orcidlink{0000-0002-1499-6377}}$,$^{2,3}$
Caroline Foster$^{\orcidlink{0000-0003-0247-1204}}$,$^{2,8}$
\newauthor
Katherine E. Harborne$^{\orcidlink{0000-0002-2043-7985}}$,$^{2,9}$
Claudia D. P. Lagos$^{\orcidlink{0000-0003-3021-8564}}$,$^{2,9}$
Di Wang$^{\orcidlink{0000-0003-3389-6264}}$,$^{1,2}$
Sabine Bellstedt$^{\orcidlink{0000-0003-4169-9738}}$,$^{9}$
\newauthor
Joss Bland-Hawthorn$^{\orcidlink{0000-0001-7516-4016}}$,$^{1,2}$
Matthew Colless$^{\orcidlink{0000-0001-9552-8075}}$,$^{2,3,10}$
Francesco D'Eugenio$^{\orcidlink{0000-0003-2388-8172}}$,$^{11,12}$
Kathryn Grasha\thanks{ARC DECRA Fellow}$^{\orcidlink{0000-0002-3247-5321}}$,$^{2,3}$
\newauthor
Yingjie Peng,$^{13,14}$
Giulia Santucci$^{\orcidlink{0000-0003-3283-4686}}$,$^{2,9}$
Sarah M. Sweet$^{\orcidlink{0000-0002-1576-2505}}$,$^{2,15}$
Sabine Thater$^{\orcidlink{0000-0003-1820-2041}}$,$^{16}$
Lucas M. Valenzuela$^{\orcidlink{0000-0002-7972-9675}}$,$^{17}$
\newauthor
and Bodo Ziegler$^{\orcidlink{0000-0003-2856-1080}}$$^{16}$
\\
$^{1}$Sydney Institute for Astronomy (SIfA), School of Physics, The University of Sydney, NSW 2006, Australia\\
$^{2}$ARC Centre of Excellence for All Sky Astrophysics in 3 Dimensions (ASTRO 3D)\\
$^{3}$Research School of Astronomy and Astrophysics, Australian National University, Canberra, ACT 2611, Australia\\
$^{4}$Astronomy, Astrophysics and Astrophotonics Research Centre, Macquarie University, Sydney, NSW 2109, Australia\\
$^{5}$School of Mathematical and Physical Sciences, Macquarie University, NSW 2109, Australia\\
$^{6}$Centre for Astrophysics and Supercomputing, School of Science, Swinburne University of Technology, Hawthorn, VIC 3122, Australia\\
$^{7}$Brain and Mind Centre, The University of Sydney, NSW 2006, Australia\\
$^{8}$School of Physics, University of New South Wales, Sydney, NSW 2052, Australia\\
$^{9}$International Centre for Radio Astronomy (ICRAR), M468, The University of Western Australia, 35 Stirling Highway, Crawley, WA 6009, Australia\\
$^{10}$Sub-Department of Astrophysics, Department of Physics, University of Oxford, Denys Wilkinson Building, Keble Road, Oxford OX1 3RH, UK\\
$^{11}$Kavli Institute for Cosmology, University of Cambridge, Madingley Road, Cambridge CB3 0HA, UK\\
$^{12}$Cavendish Laboratory, University of Cambridge, 19 JJ Thomson Avenue, Cambridge CB3 0HE, UK\\
$^{13}$Department of Astronomy, School of Physics, Peking University, 5 Yiheyuan Road, Beijing 100871, People’s Republic of China\\
$^{14}$Kavli Institute for Astronomy and Astrophysics, Peking University, 5 Yiheyuan Road, Beijing 100871, People’s Republic of China\\
$^{15}$School of Mathematics and Physics, University of Queensland, St Lucia, Queensland 4072, Australia\\
$^{16}$University of Vienna, Department of Astrophysics, Türkenschanzstraße 17, 1180 Vienna\\
$^{17}$Universitäts-Sternwarte, Fakultät für Physik,  
Ludwig-Maximilians-Universität München, Scheinerstr. 1, 81679 München, Germany\\
}

\date{Accepted 2024 August 20. Received 2024 August 8; in original form 2024 May 16}

\pubyear{2024}

\begin{document}
\label{firstpage}
\pagerange{\pageref{firstpage}--\pageref{lastpage}}
\maketitle

\begin{abstract}
We measure the ionised gas velocity dispersions of star-forming galaxies in the MAGPI survey ($z\sim0.3$) and compare them with galaxies in the SAMI ($z\sim0.05$) and KROSS ($z\sim1$) surveys to investigate how the ionised gas velocity dispersion evolves. For the first time, we use a consistent method that forward models galaxy kinematics from $z=0$ to $z=1$. This method accounts for spatial substructure in emission line flux and beam smearing. We investigate the correlation between gas velocity dispersion and galaxy properties to understand the mechanisms that drive gas turbulence. We find that in both MAGPI and SAMI galaxies, the gas velocity dispersion more strongly correlates with the star-formation rate surface density ($\Sigma_{\rm SFR}$) than with a variety of other physical properties, and the average gas velocity dispersion is similar, at the same $\Sigma_{\rm SFR}$, for SAMI, MAGPI and KROSS galaxies. The results indicate that mechanisms related to $\Sigma_{\rm SFR}$ could be the dominant driver of gas turbulence from $z\sim1$ to $z\sim0$, for example, stellar feedback and/or gravitational instability. The gas velocity dispersion of MAGPI galaxies is also correlated with the non-rotational motion of the gas, illustrating that in addition to star-formation feedback, gas transportation and accretion may also contribute to the gas velocity dispersion for galaxies at $z\sim 0.3$. KROSS galaxies only have a moderate correlation between gas velocity dispersion and $\Sigma_{\rm SFR}$ and a higher scatter of gas velocity dispersion with respect to $\Sigma_{\rm SFR}$, in agreement with the suggestion that {\secRevise other mechanisms, such as gas transportation and accretion, are relatively more important at higher redshift galaxies}.
\end{abstract}



\section{INTRODUCTION}

The evolution of the bulk gas motions within galaxies with cosmic time is governed by {\secRevise the} complex interplay of different internal and external physical processes such as gravity, feedback, cooling, gas accretion and galaxy interactions \citep[e.g.][]{Bower:2006,Dekel:2009,Bouche:2010,Dave:2012,Genel:2012,Lilly:2013,Gabor:2014,Schaye:2015,Sillero:2017,Bird:2021,Forbes:2023}. Galaxy kinematic properties are important tools to help us understand the evolution of galaxies from {\secRevise the} early Universe to the present time and to constrain galaxy evolution models \citep[e.g. review by][]{Cappellari:2016}. 

{\teamRevise The velocity dispersion of ionised gas within disk galaxies} decreases with cosmic time \citep{Kassin:2012,Wisnioski:2015,Johnson:2018,Ubler:2019}, from $\sim 50$\,km\,s$^{-1}$ in galaxies at $z > 1$ to $\sim 20$\,km\,s$^{-1}$ in the local universe \citep{Genzel:2006,Epinat:2008,Law:2009,Green:2014,Moissev:2015,Varidel:2020,Law:2022}. The thermal broadening of {\teamRevise star-forming gas} traced by the H$\alpha$ emission line due to a characteristic temperature of $10^4$K is estimated to be 11.7 km s$^{-1}$ \citep{Bland-Hawthorn:2024}\footnote{This value assumes a mean molecular weight made up of 74\% H, 24\% He and 2\% for all other metals, typical of the modern day ISM.}.
However, the ionised gas velocity dispersion observed at all epochs {\secRevise is} greater than this value.
The gas velocity dispersion above the thermal contribution must arise from other local or global processes which induce turbulence in the gas. Two questions thus arise: What are the main drivers of turbulence in ionised gas and how do the physical processes that induce turbulence evolve with cosmic time?

The main drivers of high gas velocity dispersion over cosmic time are not well constrained. Disk galaxies in a state of equilibrium must maintain a balance between the energy loss by turbulent dissipation and energy injection from other local and global mechanisms \citep{Krumholz:2018}. Global or external mechanisms that may induce higher turbulence include gas accretion \citep{Klessen:2010,Aumer:2010} and minor mergers \citep{Bournaud:2009}. Possible internal mechanisms of gas turbulence include star-formation feedback \citep[e.g.][]{Green:2010,Faucher-Giguere:2013,Green:2014,Moiseev:2015,Bacchini:2020,Egorov:2023}, gravitational instability \citep[e.g.][]{Krumholz:2016,Krumholz:2018,Molina:2020} and interactions between different galaxy components \citep{Aumer:2010,Oliva-Altamirano:2018}.

Star-formation feedback is widely considered to be a significant source of gas turbulence, both theoretically \citep{Ostriker:2010,Ostriker:2011,Faucher-Giguere:2013,Krumholz:2016,Krumholz:2018,Bacchini:2020} and in simulations \citep[e.g.][]{Shetty:2012,Rathjen:2023,Hung:2019}. Stellar feedback injects energy and momentum into the interstellar medium through supernovae, stellar winds, radiation pressure and ionising radiation. A correlation between star formation rate (SFR) and gas velocity dispersion has been found in many observational studies \citep[e.g.][]{Green:2010,Green:2014,Zhou:2017,Varidel:2020,Law:2022}, providing evidence for the role of stellar feedback in increasing the local turbulence in disks and therefore the measured disk velocity dispersion.

However, the degree of importance of stellar feedback in driving gas turbulence is debated. Some studies proposed that star formation alone can explain the gas velocity dispersion we observe \citep{Bacchini:2020, Faucher-Giguere:2013,Egorov:2023}, while other studies suggested that stellar feedback can only drive part of the gas velocity dispersion, thus other mechanisms are needed \citep{Zhou:2017,Bik:2022,Forbes:2023,Kim&Ostriker&kim:2013,Dib:2006,Shetty:2012}. Gravitational instability \citep{Krumholz:2016,Krumholz:2018}, gas accretion \citep{Klessen:2010} and galaxy interaction \citep{Bournaud:2009} have been suggested as possible sources of the additional gas turbulence. 

The importance of different mechanisms may vary with redshift and environment. Gas accretion at high redshift {\secRevise was} more efficient in converting bulk kinetic energy into turbulent kinetic energy than at low redshifts \citep{Klessen:2010,Ginzburg:2022}. {\secRevise A} simulation study by \citet{Jimenez:2022} showed that stellar feedback is more important for galaxies in lower halo masses and at lower redshift. \citet{Krumholz:2018} predicts {\secRevise that the} relation between gas velocity dispersion and SFR {\secRevise changes} with the fraction of the interstellar medium that is in {\secRevise the} star-forming molecular phase, which in turn depends on redshift.

The measurement of intrinsic gas velocity dispersion is challenging. Ionised gas velocity dispersion is normally measured based on the width of emission lines, such as the H$\alpha$ emission line. However, the measurement of line width within a dynamic system like a disk is complicated by observational and instrumental factors including the line spread function (LSF), spaxel grid, and the size and shape of the point spread function (PSF) \citep{Davies:2011,Oh:2022}. The beam smearing caused by the Earth’s atmosphere blurs light between neighbouring spaxels. If there is a velocity difference between neighbouring spaxels, the beam smearing artificially broadens emission lines, thus increasing the apparent gas velocity dispersion. The beam smearing is the strongest at the centre of galaxies, where the velocity gradient is the largest.

Several different methods have been applied to minimise or account for {\secRevise the} beam smearing. \citet{Johnson:2018} applied {\secRevise a} correction to gas velocity dispersion based on galaxy size, seeing PSF and velocity gradient. $^\mathrm{3D}$BAROLO \citep{DiTeodoro:2015}, GalPaK$^\mathrm{3D}$ \citep{Bouche:2015} and GBKFIT \citep{Bekiaris:2016} use model-fitting {\secRevise techniques} to model the flux and kinematics of galaxy three-dimensional data cubes, including beam smearing effects. $^\mathrm{3D}$BAROLO uses a tilted-ring model that assumes {\secRevise a galaxy is a thin} disc and consisting of concentric rings with different rotational velocity and flux. GalPaK$^\mathrm{3D}$ and GBKFIT assume parametric radial flux distributions and velocity profiles. 

In this study, we use the code \textsc{blobby3d} \citep{Varidel:2019} which employs a forward modelling technique that models the gas kinematics and substructures simultaneously. The advantage of \textsc{blobby3d} is that it does not assume any given gas structure, but decomposes the gas distribution into {\secRevise a number of} Gaussian `blobs'. The number of blobs is flexible {\secRevise and} depends on the complexity of the observed gas substructure. \textsc{blobby3d} has been shown to recover the intrinsic gas velocity dispersion better than other modelling techniques where there is sufficient signal-to-noise (S/N). More constrained techniques, such as $^\mathrm{3D}$BAROLO and GalPaK$^\mathrm{3D}$, may work better where S/N is poor.

Different surveys have employed various measurement techniques to measure the intrinsic gas velocity dispersion in galaxies across cosmic epochs. For example, \citet{Ubler:2019} measured the gas velocity dispersion of galaxies at $z>0.6$ by forward-modelling the one-dimensional velocity and velocity dispersion profile. \citet{Law:2022} corrected for the effects of beam smearing using estimates derived from a three-dimensional model that is constructed based on the observed velocity field of each galaxy for low redshift galaxies ($z\sim0.05$). The systematic difference between the surveys makes it difficult to constrain the evolution of gas velocity dispersion.

The most rapid change in gas velocity dispersion seems to occur from $z\sim1$ to $z\sim0$ \citep{Wisnioski:2015,Ubler:2019,Kassin:2012}. Previous studies measured the ionised gas velocity dispersion for many local galaxies from the Integral Field Spectroscopy (IFS) survey such as SAMI \citep{Varidel:2020} and MaNGA \citep{Yu:2019,Law:2022} and high redshift galaxies at $z>0.6$ such as KMOS$^\mathrm{3D}$ \citep{Ubler:2019}. {\secRevise \citet{Kassin:2012} used the slit-based DEEP2 survey to investigate the evolution of gas velocity dispersion of galaxies at $0.2<z<1.2$.} However, the gas velocity dispersion from IFS data between $0<z<1$, at which most of the change of gas velocity dispersion happens, has not been investigated in a self-consistent manner. 

In this study, we {\secRevise will} measure the ionised gas velocity dispersion of galaxies from the Middle Ages Galaxy Properties with Integral Field Spectroscopy (MAGPI) survey \citep{Foster:2021} at $z\sim 0.3$, to fill the gap of the IFS measurement of gas velocity dispersion at intermediate redshift, and {\secRevise to link SAMI galaxies \citep{Bryant:2015} at $z \sim 0.05$ and KROSS galaxies \citep{Stott:2016} at $z \sim 1$. We measure the gas velocity dispersion from these three surveys in a consistent manner, enabling a detailed exploration of how gas kinematics evolved over the past 8 billion years.} In Section \ref{sec:data}, we describe {\secRevise the} MAGPI survey, data reduction and our sample selection. In Section \ref{sec:method}, we describe the methods used to measure the gas velocity dispersion. In Section \ref{sec:results}, we present our results for MAGPI gas velocity dispersions, investigate their relations with galaxy properties and compare them with other {\secRevise surveys}. In Section \ref{sec:discussion}, we discuss the evolution of gas velocity dispersion and drivers of gas turbulence. In Section \ref{sec:conclusion}, we summarise our results. 

Throughout this paper, we adopt the concordance cosmology ($\Omega_{\Lambda} = 0.7$, $\Omega_{m} = 0.3$, $H_0 = 70\: \mathrm{km\: s^{-1}\: Mpc^{-1}}$).

\section{DATA}
\label{sec:data}

\subsection{The MAGPI Survey}

Data used in this work are drawn from the MAGPI survey \citep{Foster:2021}. {\secRevise MAGPI makes use of the ground layer adaptive optics (GLAO) aided, wide-field optical integral field spectrograph, the Multi Unit Spectroscopic Explorer \citep[MUSE;][]{Bacon:2010}, mounted at the European Southern Observatory Very Large Telescope (VLT).} MUSE has a 1 $\times$ 1 arcmin$^{2}$ field-of-view (FOV) sampled at 0.2 $\times$ 0.2 arcsec$^{2}$. The typical full width at half maximum (FWHM) of MUSE PSF is $\sim$0.6\,arcsec. \referee{ The spectroscopic observations from the MAGPI survey cover the wavelength ranges 4700-9351 \AA\ with a spectral resolution of $R = 2728$ {\firstRevise at 7025 \AA}.}

\referee{The MAGPI survey targets 60 massive central galaxies (primary galaxies) and their neighbouring galaxies (secondary galaxies) within the MUSE FOV. The targets are selected from G12, G15 and G23 fields in the Galaxy And Mass Assembly (GAMA) survey \citep{Driver:2011}.} \referee{At the time of writing, the MAGPI survey has completed the observation of 48 fields, among which 35 fields have been fully reduced with relevant data products.}. \referee{There are 780 passive and star-forming galaxies in the redshift range of interest ($0.22<z<0.42$), of which 180 have H$\alpha$-based SFR measurement.}

\subsubsection{Galaxy properties} \label{sec:magpi galaxy properties}

We use the Galaxy Integral Field Unit (IFU) Spectroscopy Tool \citep[\textsc{GIST};][]{Bittner:2019} to fit emission lines and continuum spectra in MAGPI galaxies (Battisti et al. in prep). \textsc{GIST} is a Python wrapper designed to run the Penalized PiXel-Fitting \citep[pPXF;][]{Cappellari:2004,Cappellari:2017} method on IFS data. \textsc{GIST} can extract stellar kinematics, conduct emission-line analysis, and deduce stellar population properties through spectral fitting. We use the H$\alpha$ flux map in \textsc{GIST} to calculate the half-light radius ($R_\mathrm{e}$) of MAGPI galaxies as the tracer of star-forming $R_\mathrm{e}$, inside which half the stars are currently forming. However, we note that H$\alpha$ $R_\mathrm{e}$ can be different to star-forming $R_\mathrm{e}$ if extinction significantly varies across the galaxy. At low surface brightness H$\beta$ may not have sufficient S/N to get a reliable extinction correction, so we use {\teamRevise non-extinction corrected} H$\alpha$ $R_\mathrm{e}$. We generate {\secRevise the} inclination corrected curve of growth of H$\alpha$ flux as a function of radius and {\secRevise define the radius enclosing 50\% of the flux as $R_\mathrm{e}$, to clarify that this is effectively the major axis $R_\mathrm{e}$, not circularly averaged.} {\teamRevise Throughout this paper, $R_\mathrm{e}$ represents the half-light radius of H$\alpha$ flux unless specified otherwise.}

\textsc{blobby3d} requires stellar continuum-subtracted data. We subtract the stellar continuum from the MUSE spectrum using the stellar continuum fitting products from GIST. The stellar continuum fits were done on a spaxel level. Initial estimates of the stellar continuum were performed with the emission line masked, and those initial estimates were then used to fit the stellar continuum and emission line using templates from flexible stellar population synthesis \citep[FSPS;][]{Conroy:2009,Conroy:2010a,Conroy:2010b} and the multiplicative Legendre polynomial is set to 12 (Battisti et al. in prep). 

We use the Baldwin, Phillips \& Terlevich \citep[BPT;][]{baldwin:1981,Kewley:2001,Kauffmann:2003} ionization diagnostics to identify star-forming (SF), active galactic nucleus (AGN) and composite spaxels in galaxies and to classify galaxies. We use the line ratios [NII]$\lambda6584$/H$\alpha$ and [OIII]$\lambda5007$/H$\beta$, {\secRevise including only spaxels that have S/N>3 in all emission lines}. The flux and error of these four emission lines are estimated {\secRevise using} GIST \citep[][Battisti et al. in prep]{Bittner:2019} from 50 Monte Carlo realisations of the input data. We remove galaxies that are dominated by AGN and/or composite spaxels ($\geq$50\%) from our sample. However, we keep the galaxies with AGN and/or composite centre (>10 spaxels) and star-forming disc ($\geq$50 SF spaxels), and identified them as AGN-host galaxies. We mask the AGN and/or composite centres and keep the SF discs for these galaxies. Galaxies that are dominated by SF spaxels are classified as SF galaxies. Some SF galaxies may have a few spaxels at the edge of galaxies classified as composite spaxels due to the low S/N of the [NII]$\lambda6584$, [OIII]$\lambda5007$ or H$\beta$ emission lines.

{\secRevise SFRs} are estimated using the dust-corrected H$\alpha$ luminosity \citep{Kennicutt:1998}. {\firstRevise H$\alpha$ flux is corrected for extinction using the Balmer decrement on a spaxel by spaxel basis. The extinction curve from \citet{Fitzpatrick:2019} is adopted with $R_V = 3.1$. The SFRs are first calculated on a spaxel-by-spaxel basis. The integrated SFRs are then calculated from a sum of the SFRs of all SF spaxels to avoid contamination from AGN (see Mun et al. \citeyear{Mun:2024} for more details).} The fraction of SFR missed due to ignoring the AGN/composite spaxels is small. The upper limit on the flux missed (assuming all H$\alpha$ flux is from star formation) is on average only 25\% (0.12 dex) for AGN-host galaxies. {\secRevise The SFR surface density ($\Sigma_{\rm SFR}$) is calculated as $\Sigma_{\rm SFR} = \mathrm{SFR}/(2\pi R_\mathrm{e}^2)$, in which $R_\mathrm{e}$ from H$\alpha$ is used.}

{\firstRevise The stellar masses are calculated using \textsc{ProSpect} \citep{Robotham:2020}, a package used to analyse star formation history and spectral energy distributions (SED). {\secRevise Photometry over the $ugriZYJHK$ bands from GAMA imaging is used to perform \textsc{ProSpect} SED fitting, assuming a \citet{Chabrier:2003} stellar initial mass function (IMF).} The stellar population templates by \citet{Bruzual:2003} are adopted and the dust attenuation law by \citet{Charlot:2000} is applied to correct stellar light attenuation by dust, as outlined in \citet{Bellstedt:2020b}. The stellar mass surface density ($\Sigma_{M_*}$) is calculated as $\Sigma_{M_*} = M_*/(2\pi R_\mathrm{e}^2)$.} 

\subsubsection{PSF and LSF}

The spatial PSF for {\secRevise the} MAGPI survey {\secRevise is reconstructed from} information generated by the adaptive optics system \citep{Fusco:2020}. The advantage of the reconstructed PSF is that it can be obtained for any given time and field, instead of strongly depending on the availability of bright stars. {\firstRevise It agrees well with an in-field PSF estimate when a bright star is present (Mendel et al., in prep).}

The PSF convolution kernel in \textsc{blobby3d} is a sum of multiple 2D concentric circular \referee{Gaussian profiles}. {\secRevise We fit the reconstructed PSF with the two-component Gaussian model, which is enough to} adequately model the PSF profile. We also test a PSF model using three \referee{Gaussian profiles}, however this results in just a 1\% change in the measured velocity dispersions (i.e., it does not significantly affect our results), but requires 30\% more CPU time in running \textsc{blobby3d}.

The spectral LSF is measured from fitting \referee{Gaussian profiles} to isolated bright skylines, both for a collapsed spectrum of the full field and for every spaxel individually \citep[][Mendel et al. in prep]{Bacon:2017}. The spatial variation of the MUSE LSF is small, {\secRevise so we use the LSF} derived from an average spectrum of the full field. \referee{The variation of instrumental resolution ($\sigma_{\mathrm{LSF}}$) between different fields is $0.135 \pm 0.006$ km s$^{-1}$.} When {\firstRevise we} build the \textsc{blobby3d} model for different galaxies, we use the $\sigma_{\mathrm{LSF}}$ based on their field and redshift.

\subsubsection{Sample selection}
\label{sec:sample selection}

In our preliminary selection, we choose galaxies {\firstRevise that have more than 50 spaxels with H$\alpha$ S/N$>$3}. Galaxies with fewer spaxels have less robust resolved kinematics due to beam smearing. Given the typical PSF of MAGPI data (FWHM $\sim$0.6~arcsec), 50 pixels correspond to $\sim$7 independent resolution elements within the galaxy. {\firstRevise The galaxies chosen have redshifts between 0.22 and 0.42.} We remove galaxies with clearly disturbed gas kinematics and galaxies {\secRevise with} inclination greater than 60\textdegree. The intrinsic gas velocity dispersion of high-inclination galaxies is hard to constrain due to the beam smearing \citep{Varidel:2019}. Furthermore, \textsc{blobby3d} assumes galaxies have {\secRevise a} thin disc and so is not able to model very edge-on galaxies, as the thickness of the galaxy disc becomes prominent for edge-on galaxies so the thin disc model does not work well. {\secRevise There are 110 galaxies in our preliminary sample after applying the above cuts.} 

\subsection{The SAMI Galaxy Survey}

{\secRevise The Sydney–AAO Multi-object Integral field spectrograph \citep[SAMI;][]{Croom:2012} uses 13 hexabundles \citep{BlandHawthorn:2011} over 1\textdegree \, diameter field-of-view, mounted on the Anglo-Australian Telescope.} The SAMI galaxy survey \citep{Bryant:2015,Scott:2018} collected spatially resolved spectra for 3068 galaxies at $0.004<z<0.115$ {\secRevise with a broad range in galaxy stellar mass ($\mathrm{log}(M_*/M_\odot)=10^8$ to $10^{12}$).} The integral field units were hexabundles \citep{BlandHawthorn:2011} that contained 61 fibres that cover 15 arcsec diameter regions on the sky. The wavelength range of SAMI is 3750-5750 \AA\ ($R=1808$) and 6300-7400 \AA\ ($R=4304$) for the blue and red arms respectively. Stellar masses are estimated from the rest-frame $i$-band absolute magnitude and $g-i$ colour, following the method of \citet{Taylor:2011} and assuming a \citet{Chabrier:2003} IMF.

\citet{Varidel:2020} used \textsc{blobby3d} to measure the ionized gas velocity dispersion of {\secRevise 383 star-forming} SAMI galaxies. {\secRevise They removed AGN and low-ionization nuclear emission-line region (LINER) galaxies from their sample. They removed galaxies with an inclination greater than 60\textdegree. They also removed mergers or galaxies with clearly disturbed gas kinematics. They applied a mask to spaxels with H$\alpha$ S/N$<3$ and retained galaxies that had at least 300 unmasked spaxels.} \referee{They rounded the gas velocity dispersion to the nearest whole number.}

Given that both \citet{Varidel:2020} and this work employ spatially resolved spectroscopic observational data and utilize the same dispersion measurement methods, we adopt the gas velocity dispersions measured by \citet{Varidel:2020} as a source of low-redshift dispersion measurements for comparison. For the SAMI galaxies, we use $R_\mathrm{e}$ measured from the H$\alpha$ emission line \citep[SAMI DR3;][]{Croom:2021} as discussed for MAGPI galaxies in Section \ref{sec:magpi galaxy properties}, and dust-corrected H$\alpha$-based SFRs \citep{Kennicutt:1998}, using the dust extinction law of \citet{Cardelli:1989}, for better comparisons. We note that \citet{Varidel:2020} used $R_\mathrm{e}$ measured {\secRevise using {\firstRevise $r$-band} images} \citep{Bryant:2015} and SFRs measured by MAGPHYS \citep{daCunha:2008} using full spectral energy distribution fits to 21 photometric bands spanning the UV, optical and far-infrared.

\subsection{KROSS}

The KMOS Redshift One Spectroscopic Survey \citep[KROSS;][]{Stott:2016} observed 795 star-forming galaxies at $z$=0.8--1.0 using the $K$-band Multi Object Spectrograph \citep[KMOS;][]{Sharples:2013} on the VLT. Each IFU of KMOS covers a ${2.8}''\times{2.8}''$ field with ${0.2}''\times{0.2}''$ spatial pixels. The H$\alpha$ line in this redshift range is observed within the $Y J$-band grating with spectral resolution $R=3400$. We use $R_\mathrm{e}$ measured from the H$\alpha$ flux.

{\secRevise Varidel et al. (in prep)\footnote{This work has been published in PhD thesis: https://hdl.handle.net/2123/25670} measure the gas velocity dispersion of 315 KROSS galaxies using \textsc{blobby3d}. They remove AGN galaxies and highly-inclined galaxies ($i>60$\textdegree) and galaxies where the mode of the gas velocity dispersion in the posterior distribution is smaller than the thermal contribution of the gas, 9\,km\,s$^{-1}$ (which might be caused by the variation in spectral resolution on KMOS). They also removed major mergers, i.e. galaxies with clearly disturbed kinematics. This leaves 193 galaxies for further analysis.}

We use the redshift, SFR, stellar mass, and inclination angle from the publicly available KROSS catalogue \citep{Harrison:2017}. The SFRs are calculated using the integrated H$\alpha$ luminosity with dust extinction applied \citep{Wuyts:2011}. Stellar masses are estimated by scaling the $H$-band AB magnitude by a constant mass-to-light ratio for all galaxies, assuming a \citet{Chabrier:2003} IMF. \citet{Harrison:2017} adopted the median mass-to-light ratio for the KROSS sample, $\mathrm{\gamma}_H = 0.2$.

\begin{figure*}
    
    \includegraphics [width=0.95\textwidth]{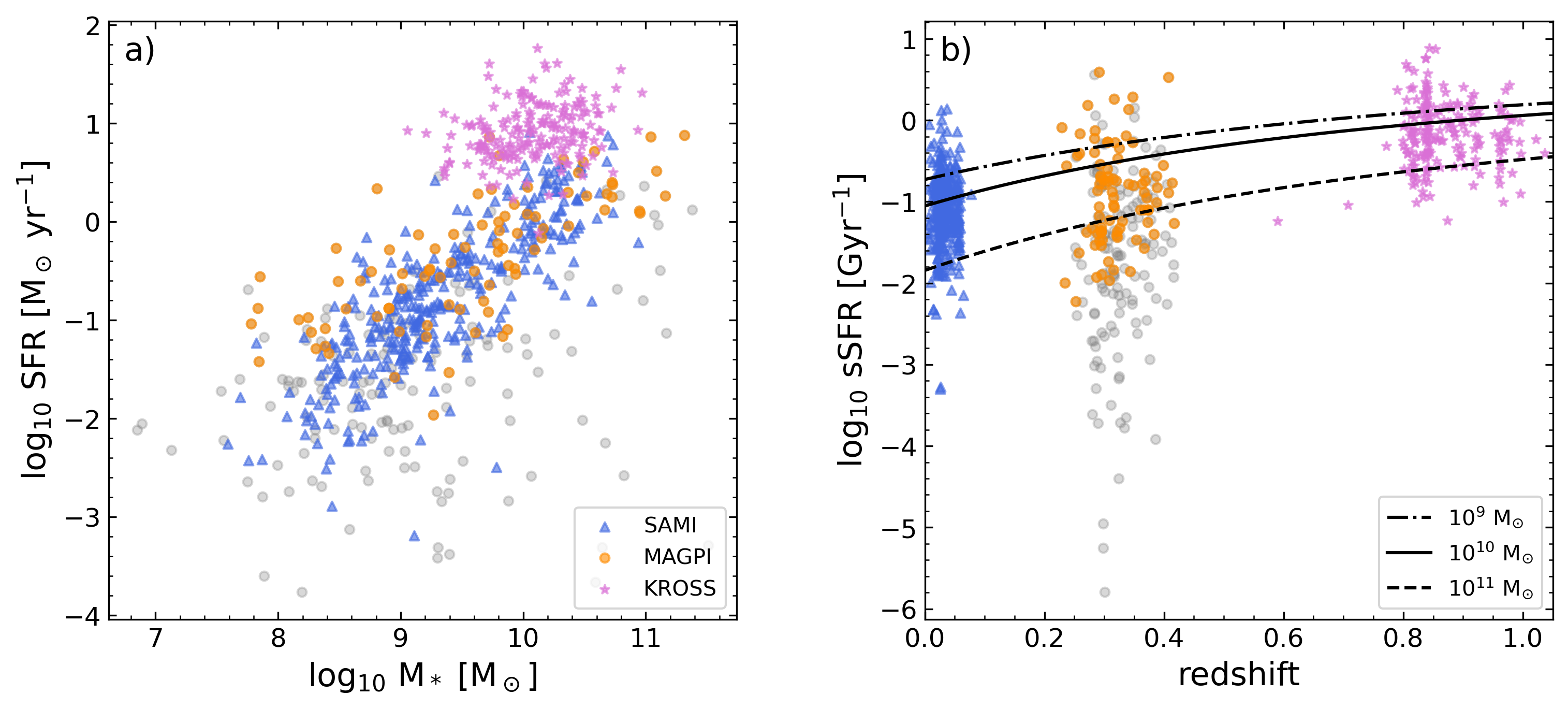}
    \caption{a) The SFR-M$_*$ relation for SAMI (blue), MAGPI (orange) and KROSS (purple) galaxies in our sample. MAGPI galaxies that have H$\alpha$-based SFR measurement but that are not in our final sample are shown in grey. b)  The sSFR-redshift relation for galaxies in our sample. The black lines show the evolution of sSFR as a function of redshift drawn from \citet{Popesso:2023} at three stellar mass values: 10$^{9}$ (dot-dashed line), $10^{10}$ (solid line) and $10^{11}$ M$_{\odot}$ (dashed line).}
    \label{fig:SFR_mass_redshift}
\end{figure*}

\referee{We show the SFR-M$_*$ relation for SAMI, MAGPI and KROSS galaxies in our sample in Figure \ref{fig:SFR_mass_redshift}(a). This shows that galaxies in the MAGPI sample have similar mass and SFR distribution, while the galaxies in the KROSS sample have higher mass and SFR. The MAGPI galaxies that have H$\alpha$-based SFR measurement but that are not in our final sample are shown as grey circles. In Figure \ref{fig:SFR_mass_redshift}(b), we show the sSFR as a function of redshift for galaxies in the surveys. The evolution of sSFR of the galaxies in our sample aligns with the result in \citet{Popesso:2023}, which captured the evolution of the star formation main sequence (SFMS).} 

\section{METHODS}
\label{sec:method}

\subsection{Modeling the gas disc kinematics}

We use the forward-modelling technique, \textsc{blobby3d}\footnote{https://github.com/SpaceOdyssey/blobby3d} \citep{Varidel:2019}, to infer the {\teamRevise spatial distribution and kinematics of the gas} simultaneously. \textsc{blobby3d} is a Bayesian inference {\secRevise code} for gas disc kinematics, using a Gaussian mixture model {\teamRevise for the spatial distribution of the gas but assuming a thin regularly rotating galaxy}. \textsc{blobby3d} does not impose a specific spatial gas structure, but instead decomposes the gas distribution into a sum of positive-definite Gaussian basis functions (blobs), which gives it more flexibility in modelling the gas distribution. \textsc{blobby3d} constructs a 3D (position-position-wavelength) cube that is convolved by the PSF in the spatial direction and the LSF in the wavelength direction. The convolved model is compared to the observed data cube. {\firstRevise \textsc{blobby3d} uses a diffusive nested sampling algorithm \citep{Skilling:2004}, \textsc{DNest4} \citep{Brewer:2011,Brewer:2018}, to run Markov Chain Monte Carlo (MCMC) chains to explore the posterior distribution. The advantage of \textsc{DNest4} over other sampling algorithms is in its application to high dimensional parameter spaces and multimodal distributions, which is essential to model the gas distribution and kinematics of galaxies. \textsc{DNest4} allows \textsc{blobby3d} to jump to different models with varying numbers of blobs. For all parameters fitted in \textsc{blobby3d}, see more details in Table 1 of \citet{Varidel:2019}. The {\secRevise results} we use in this work are the median value over the gas velocity dispersion in the posterior samples. }

\begin{figure*}
    \centering
    \includegraphics [width=0.95\textwidth]{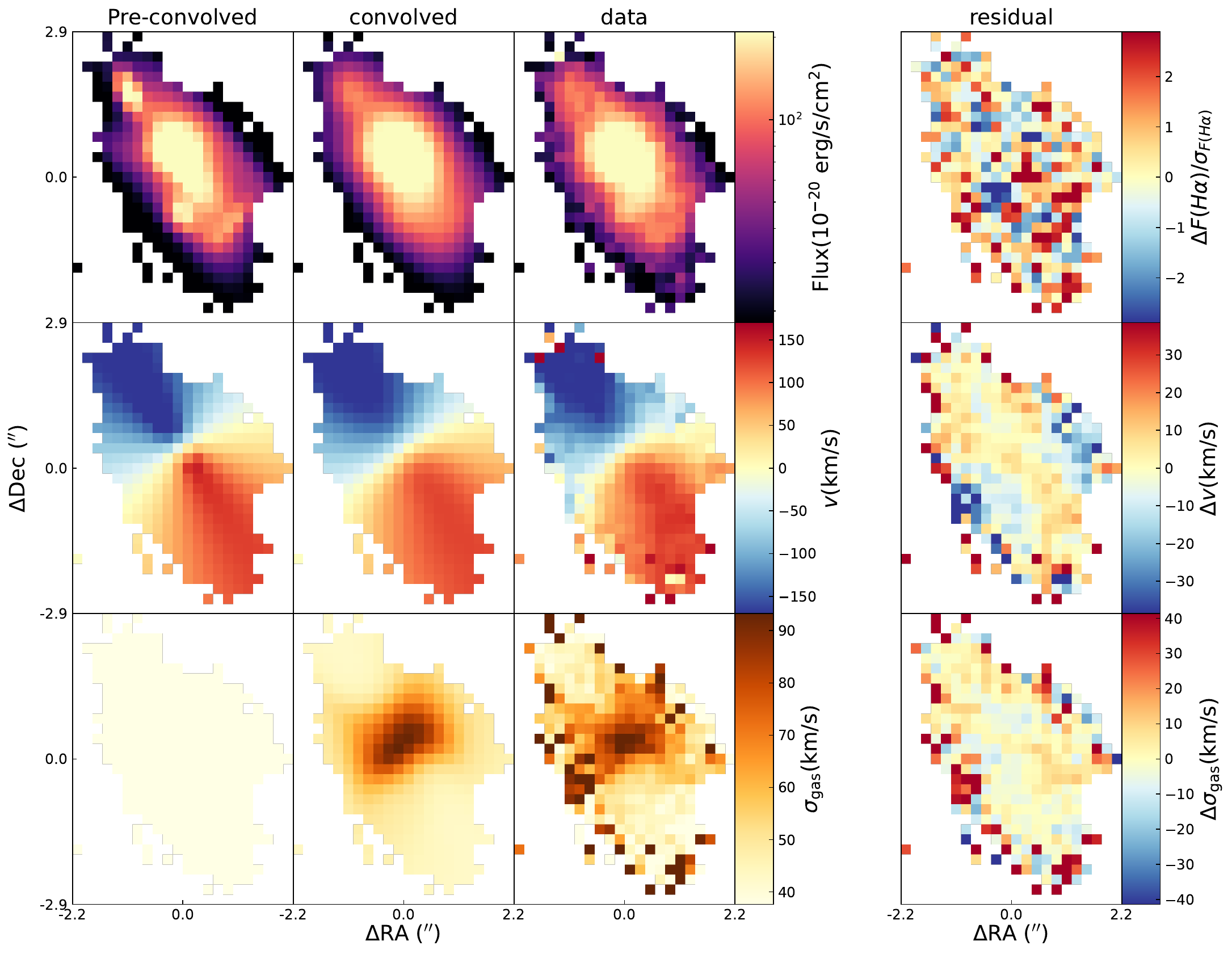}
    \caption{The 2D maps of the \textsc{blobby3d} results for MAGPI galaxy 1506106169. The first column shows the model made by \textsc{blobby3d}; the second column shows the model convolved with the PSF and LSF of the data; the third column shows the results of the single Gaussian fit to the observed datacube; the fourth column shows the residuals between the data and the convolved model. The first row shows the flux distribution of the H$\alpha$ emission line; the second row shows the velocity map; the third row shows the velocity dispersion map.}
    \label{fig:b3d_output}
\end{figure*}

{\secRevise We measure the ionised gas velocity dispersion of MAGPI galaxies using the H$\alpha$ emission line. An example set of 2D maps, showing the results of using \textsc{blobby3d} on a MAGPI galaxy 1506106169, is displayed in Figure \ref{fig:b3d_output}. This galaxy is a median-size galaxy in our sample, which has 329 spaxels with H$\alpha$ S/N greater than 3.}

\subsubsection{Velocity profile}

\textsc{blobby3d} assumes a continuous velocity profile {\firstRevise for the} galaxy, based on an empirical model proposed by \citet{Courteau:1997}
\begin{equation} \label{eq:vel_profile}
    v\left ( r \right ) = v_{\mathrm{c}}\frac{\left ( 1+r_\mathrm{t}/r \right )^{\beta }}{\left ( 1+\left ( r_\mathrm{t}/r \right )^{\gamma } \right )^{1/\gamma }}\ \textup{sin}\left ( i \right )\textup{cos}\left ( \theta  \right ) + v_{\textup{sys}}~,
\end{equation}
where $v_\mathrm{c}$ is the asymptotic velocity, $r$ is the distance to the kinematic centre, $r_\mathrm{t}$ is the turnover radius, $\beta$ is a parameter that controls the gradient of the velocity profile, $\gamma$ is a parameter that controls the sharpness of the velocity profile turnover point, $v_\mathrm{sys}$ is a systemic velocity term, $i$ is the inclination of the galaxy and $\theta$ is the polar angle in the plane of the disc. 
{\firstRevise We calculate inclinations from the axis ratio ($b/a$) of the white-light image of MAGPI {\secRevise MUSE data}, which are estimated using the tool \textsc{ProFound} \citep{Robotham:2018}, as}
\begin{equation}
    \textup{cos}\left ( i \right )=\sqrt{\frac{{}\left ( b/a \right )^{2}-q^{2}_{0}}{1-q^{2}_{0}}}~,
\end{equation}
where $q_0$ is the intrinsic axial ratio of an edge-on galaxy \citep{Weijmans:2014}; here we adopt $q_0=0.2$, following \citet{Cortese:2016}. {\teamRevise We note that the adopted value of $q_0$ may influence the estimation of $v_{\mathrm{c}}$, but the fit gas velocity dispersion is insensitive to $q_0$. The emission line broadening caused by beam smearing is related to line-of-sight velocity, which is the combination of inclination and rotational velocity. Therefore, the underestimation or overestimation of $i$ can be balanced by $v_{\mathrm{c}}$ in Eq. \ref{eq:vel_profile}.}

\subsubsection{Velocity dispersion gradient}

\textsc{blobby3d} allows flexibility to assume different forms for the velocity dispersion as a function of radius. We adopt a flat velocity dispersion profile to simplify the inference following \citet{Varidel:2020}'s analysis for SAMI galaxies. However, other surveys have shown that velocity dispersion can be radially variant \citep[e.g.][]{Genzel:2017,Yu:2019}. We therefore tested using a log-linear profile and fit all spaxels with S/N>3 in H$\alpha$. {\firstRevise We have 11 AGN-host galaxies in our preliminary sample. These galaxies tend to have a higher central velocity dispersion, which forces a steep velocity dispersion slope and leads to an under-estimate of the outer dispersion. For this reason, we mask central AGN regions based on the BPT maps \citep{Kewley:2001,Kauffmann:2003}.} {\secRevise We note that the small number of high-dispersion pixels at the edges of the velocity maps in some galaxies (as seen in Figure \ref{fig:b3d_output}) are likely spurious artifacts caused by low H$\alpha$ S/N in these regions.} This was also the case in {\secRevise the} \citet{Varidel:2020} analysis for SAMI galaxies. 

{\secRevise For comparison, we run the model fits using both a constant velocity dispersion across the entire galaxy and a gradient in the dispersion. Overall, we find flat gradients and mean velocity dispersions that do not change significantly across the galaxies in either fit. We therefore choose the simpler constant velocity dispersion model.}

\section{RESULTS} \label{sec:results}

\subsection{Velocity dispersions of MAGPI galaxies}
\label{sec:velocity dispersion measurement}

Of 110 selected galaxies in our preliminary sample, 87 have a unimodal posterior distribution in gas velocity dispersion ($\sigma_{\mathrm{gas}}$) with a peak at a realistic velocity (i.e. $>10$\,km\,s$^{-1}$). The mode of the posterior distribution of the remaining 23 galaxies is at $0\:\mathrm{km\:s^{-1}}$, which is unrealistic. This may result from an intrinsic dispersion much lower than the spectral resolution of MUSE, which \textsc{blobby3d} is not able to resolve. The galaxies that do not have reliable velocity dispersion fits have low SFR and low gas velocity dispersion, and typically have large uncertainties in gas velocity dispersion. \referee{For those galaxies without reliable fit, the 95 percentile of $\sigma_\mathrm{gas}$ of posterior samples is taken as the upper limit.}

\begin{figure}
    \centering
    \includegraphics [width=0.45\textwidth]{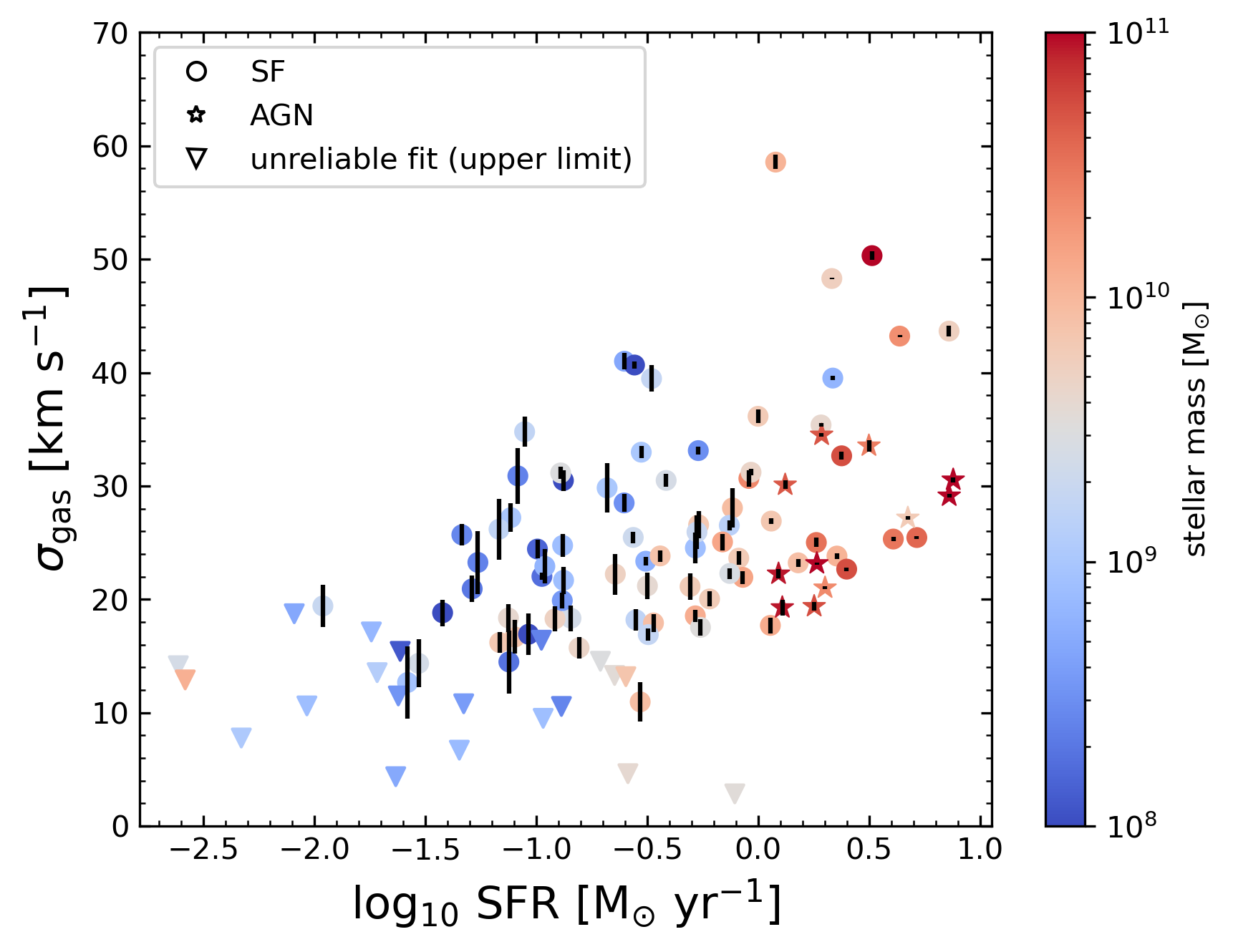}
    \caption{The intrinsic gas velocity dispersion of MAGPI galaxies measured by \textsc{blobby3d} as a function of SFR, colour-coded by stellar mass. Star-forming galaxies are shown as circles and galaxies with AGN at the centre are shown as stars. \referee{The upper limits of $\sigma_{\mathrm{gas}}$ for galaxies without reliable fits are shown as downward triangles.} The black errorbars show the uncertainties on $\sigma_{\mathrm{gas}}$.}
    \label{fig:vdispVsSFR}
\end{figure}

The average $\sigma_\mathrm{gas}$ for the 87 MAGPI galaxies that have reliable fit is $26.1\pm 8.7$\,km\,s$^{-1}$, where $\pm 8.7$\,km\,s$^{-1}$ represents the standard deviation. \referee{When we include the upper limit of $\sigma_\mathrm{gas}$ for galaxies without reliable fits, the average $\sigma_\mathrm{gas}$ for the 110 galaxies is $23.4\pm 9.8$\,km\,s$^{-1}$} Figure \ref{fig:vdispVsSFR} shows SFR versus gas velocity dispersion for 110 MAGPI galaxies, colour-coded by stellar mass. This includes 76 star-forming galaxies (circles) and 11 AGN-host galaxies (stars). The standard deviations of the posterior effective samples are shown as error bars. \referee{We show the upper limits of $\sigma_\mathrm{gas}$ for galaxies without a reliable fit (downward triangles). Note that we remove those galaxies with unreliable fits from any further analysis.}

The spectral resolution for the H$\alpha$ line of MAGPI galaxies corresponds to $\sim 38$\,km\,s$^{-1}$. {\firstRevise The minimum velocity dispersion that converges in the galaxies we selected} is 11.0\,km\,s$^{-1}$, about 1/3 of the instrumental resolution. It is difficult to measure the intrinsic velocity dispersion if it is much lower than the instrumental resolution (Wisnioski et al. in prep).

Most of the galaxies with AGN sit at high SFR in Figure \ref{fig:vdispVsSFR}. These galaxies also have larger sizes and higher stellar masses (see Section \ref{sec:correlation_sigma_properties}), which are positively correlated with SFR. {\firstRevise However, it is noteworthy that they tend to have lower dispersion than other galaxies of the same SFR.} We test whether the masking of the centres of the AGN galaxies influences our measurements by selecting 9 SF galaxies, masking their centres and then refitting for $\sigma_\mathrm{gas}$. This test shows that the differences of $\sigma_{\mathrm{gas}}$ between the centre-masked and no-mask versions are a few km\,s$^{-1}$ and the average offset is 0.4 km\,s$^{-1}$. {\secRevise This suggests that our approach of masking the central region in AGN-host galaxies does not bias the $\sigma_\mathrm{gas}$ measurements.}

Galaxies with AGN at the centre do not have higher dispersion than the galaxies {\secRevise with} similar SFR, which indicates that {\secRevise the AGN in our sample}, though they may have high gas velocity dispersion {\secRevise in the central} AGN region, do not influence the global gas velocity dispersion of the star-forming disc of their host galaxies. {\firstRevise AGN-host galaxies are closer to the average values of dispersion at fixed $\Sigma_{\mathrm{SFR}}$ {\secRevise (see Section \ref{sec:correlation_sigma_properties} below)}.} 

\begin{figure*}
    \centering
    \includegraphics [width=0.95\textwidth]{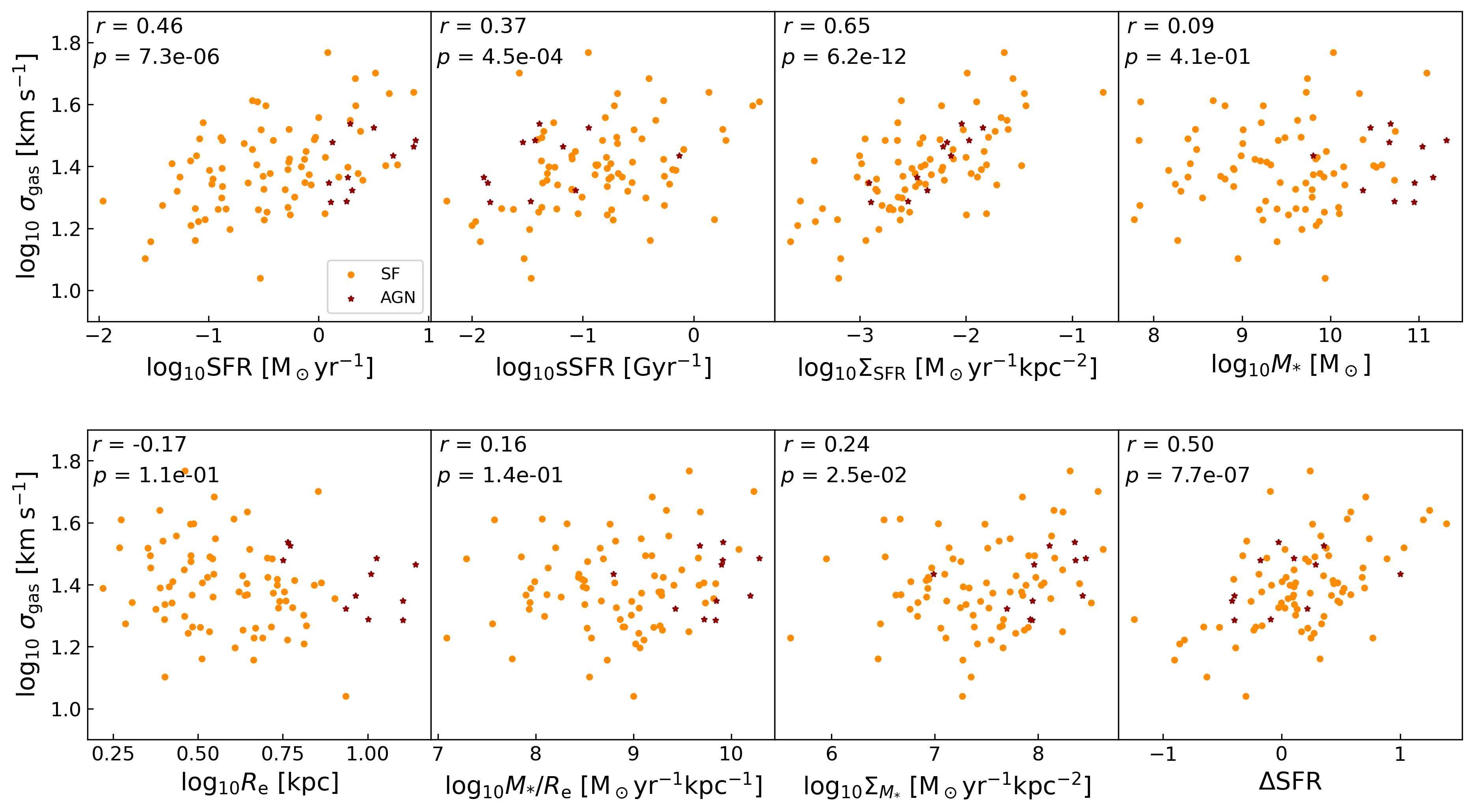}
    \caption{Intrinsic gas velocity dispersion ($\sigma_{\mathrm{gas}}$) as a function of galaxy properties for MAGPI galaxies, including star-forming rate (SFR), specific star-formation rate (sSFR), star-formation rate surface density ($\Sigma_\mathrm{SFR}$), stellar mass ($M_*$), half-light radius of H$\alpha$ flux ($R_\mathrm{e}$), a proxy for the gravitational potential ($M_*/R_\mathrm{e}$), stellar mass surface density ($\Sigma_{M_*}$), and the logarithmic offset in SFR from the star formation main sequence of MAGPI galaxies($\Delta$SFR). Pearson correlation coefficient, $r$, and $p$-value for each correlation are shown on each panel. The star-forming galaxies are shown as orange circles and the AGN-host galaxies are shown as dark red stars. We find the $\sigma_{\mathrm{gas}}$ of MAGPI galaxies has the strongest correlation with $\Sigma_\mathrm{SFR}$ ($r=0.65$). The $\sigma_{\mathrm{gas}}$ also has a strong correlation with SFR and $\Delta$SFR.} 
    \label{fig:sigma_properties}
\end{figure*}

\subsection{Correlations between velocity dispersion and other galaxy properties}
\label{sec:correlation_sigma_properties}

To investigate what factors can impact gas velocity dispersion, we calculate the Pearson correlation coefficient ($r$) between gas velocity dispersion and galaxy properties, such as total SFR, specific SFR (sSFR), $\Sigma_{\mathrm{SFR}}$, stellar mass, $R_\mathrm{e}$, a proxy for the gravitational potential ($M_*/R_\mathrm{e}$), $\Sigma_{M_*}$, and the logarithmic offset in SFR from the SFMS of MAGPI galaxies ($\Delta$SFR). The SFMS is defined as the best linear fit to the MAGPI galaxies located at $0.25 \leq z \leq 0.424$ that have SFRs measured from H$\alpha$ \citep{Mun:2024}. We plot all these correlations in Figure \ref{fig:sigma_properties}, with the Pearson correlation coefficient ($r$) and $p$-value on each plot. The Pearson correlation coefficient measures the linear correlation between two variables. The $p$-value helps determine whether the observed correlation is statistically significant. A $p$-value of 0.05 or lower is generally considered statistically significant.

We find that the gas velocity dispersion of MAGPI galaxies is strongly correlated with SFR, as already shown in Figure \ref{fig:vdispVsSFR}, with correlation coefficient $r=0.46$. {\firstRevise We look at properties derived using SFR to see if they reduce the scatter in {\secRevise the} $\sigma_{\mathrm{gas}}-\mathrm{SFR}$ relation. When SFR {\secRevise is combined} with $R_\mathrm{e}$, the $\Sigma_{\mathrm{SFR}}$ gives a stronger correlation with $\sigma_{\mathrm{gas}}$ ($r=0.65$) than SFR, while $R_\mathrm{e}$ has no significant correlation with $\sigma_{\mathrm{gas}}$ ($r=-0.17$).} This result is consistent with galaxies at $z\sim 0.05$ \citep[SAMI;][]{Varidel:2020}. This trend may arise from star-formation feedback, mainly from the supernova, which injects energy into neighbouring gas, leading to turbulence and increasing the gas velocity dispersion \citep[e.g.][]{Faucher-Giguere:2013,Krumholz:2018}. The rate of energy injection per unit area, traced by $\Sigma_{\mathrm{SFR}}$, is more correlated with gas velocity dispersion than the global SFR of galaxies, which {\secRevise indicates that the star formation on small scales within the disc may be more important than the global SFR of the galaxy.}

We also calculate the $\Sigma_{\mathrm{SFR},{i-\mathrm{band}}}$ using $R_\mathrm{e,{i-\mathrm{band}}}$ ($\Sigma_{\mathrm{SFR},{i-\mathrm{band}}} = \mathrm{SFR}/(2\pi R_{\mathrm{e},{i-\mathrm{band}}}^2)$). The $R_{\mathrm{e},{i-\mathrm{band}}}$ is the half-light radius of $i$-band image of MAGPI galaxies that {\secRevise is estimated by fitting a} single Sérsic profile using Galfit \citep{Peng:2010}. We find that the $\Sigma_{\mathrm{SFR}}$ using $R_\mathrm{e}$ of the H$\alpha$ flux gives a higher correlation coefficient ($r=0.65$) than $\Sigma_{\mathrm{SFR},{i-\mathrm{band}}}$ ($r=0.61$). {\secRevise As we might expect, $\Sigma_{\mathrm{SFR}}$ calculated by H$\alpha$ $R_\mathrm{e}$ has a tighter correlation, because it more accurately portrays the $\Sigma_{\mathrm{SFR}}$ of star-forming regions within the galaxies.}

We incorporate SFR with stellar mass in two different ways, i.e. sSFR and $\Delta$SFR. $\Delta$SFR has a slightly stronger correlation ($r=0.50$) with $\sigma_{\mathrm{gas}}$ than SFR. sSFR does not reduce the scatter in $\sigma_{\mathrm{gas}}-\mathrm{SFR}$ relation, with $r=0.37$ for sSFR. Also, we find that M$_*$ have no significant correlation with $\sigma_{\mathrm{gas}}$ ($r=0.09$). There is no significant correlation between $\sigma_\mathrm{gas}$ and $M_*/R_\mathrm{e}$ ($r=0.16$), in contrast to stellar velocity dispersion correlating with potential, which indicates that the gas velocity dispersion is more easily influenced by local power sources \citep{Oh:2022}. We note that M$_*$/$R_{\mathrm{e},{i-\mathrm{band}}}$ is a better proxy for gravitational potential, but the correlation coefficient does not change when we use M$_*$/$R_{\mathrm{e},{i-\mathrm{band}}}$. We use H$\alpha$ $R_\mathrm{e}$ for all parameters in Figure \ref{fig:sigma_properties} for consistency. $\Sigma_{M_*}$ is weakly correlated with $\sigma_{\mathrm{gas}}$ ($r=0.24$). 

\begin{figure}
    \centering
    \includegraphics [width=0.45\textwidth]{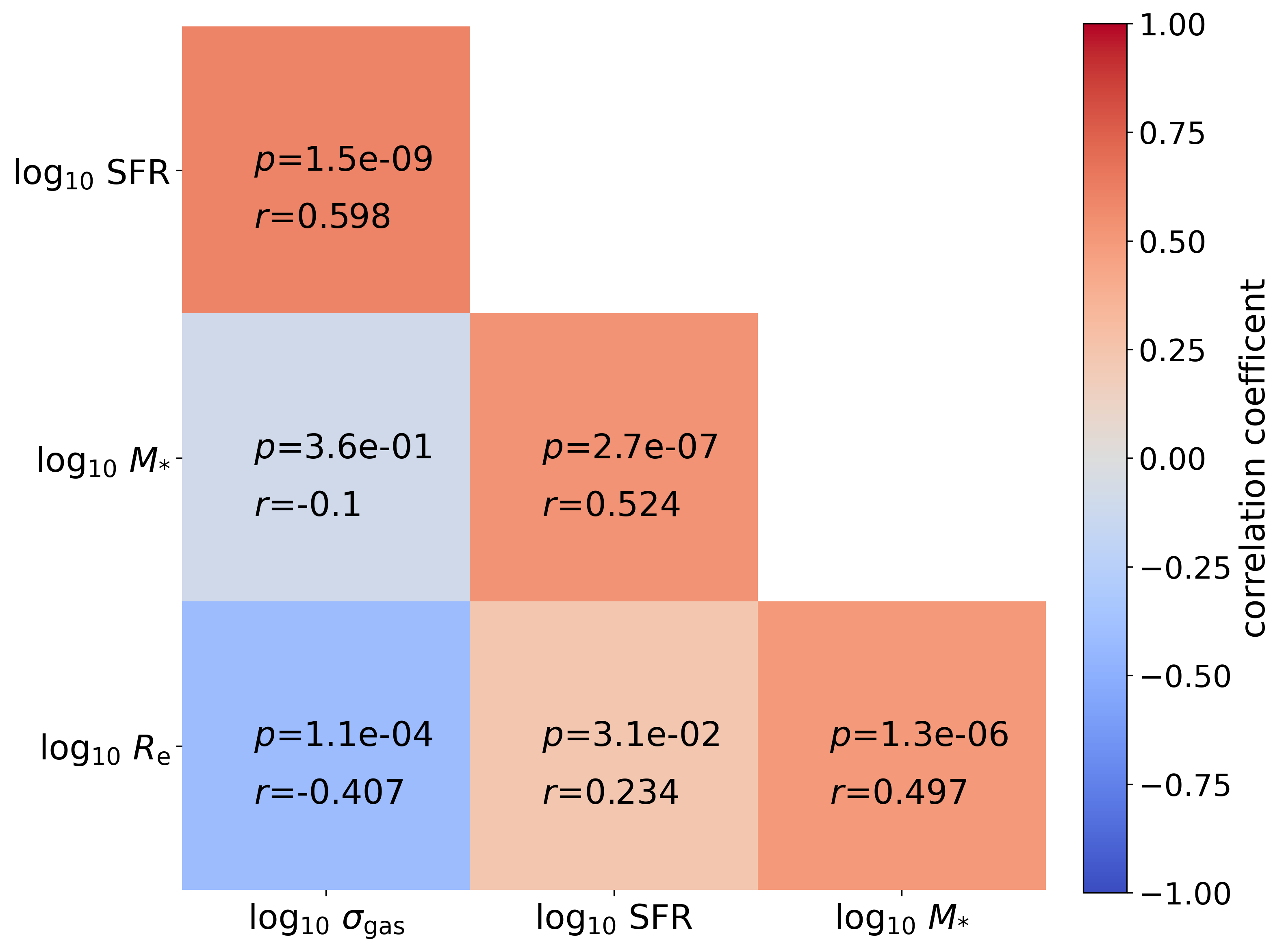}
    \caption{The partial correlations between gas velocity dispersion, SFR, $M_*$ and $R_\mathrm{e}$ for the MAGPI sample. {\secRevise The Pearson correlation coefficient ($r$) and $p$-value are given in each box, which is colour-coded by its correlation coefficient.} The partial correlation shows that SFR still has a strong correlation with gas velocity dispersion ($r= 0.598$) when we remove the effects from $M_*$ and $R_\mathrm{e}$. $R_\mathrm{e}$ has a negative partial correlation with $\sigma_{\mathrm{gas}}$ ($r = -0.407$), indicating that SFR and galaxy size have a combined effect on gas velocity dispersion.}
    \label{fig:partial_corr}
\end{figure}

From the galaxy SFMS, we know that galaxies with higher mass tend to have higher SFR. To disentangle the correlation between these parameters, we {\secRevise carry out a} partial correlation analysis as shown in Figure \ref{fig:partial_corr} using \textsc{Pingouin} \citep{Vallat2018}. Partial correlation measures the correlation between two variables, with the effect of other variables removed. The partial correlations between $\sigma_{\mathrm{gas}}$, SFR, $M_*$ and $R_\mathrm{e}$ show that SFR has the strongest correlation with $\sigma_{\mathrm{gas}}$ ($r= 0.598$) when we take $M_*$ and $R_\mathrm{e}$ into account. $M_*$ has no significant partial correlation with $\sigma_{\mathrm{gas}}$ ($r = -0.1$). Though no significant correlation has been found between $\sigma_{\mathrm{gas}}$ and $R_\mathrm{e}$ in Figure \ref{fig:sigma_properties}, $R_\mathrm{e}$ has a negative partial correlation with $\sigma_{\mathrm{gas}}$ ($r = -0.407$), indicating that SFR and galaxy size have a combined effect on gas velocity dispersion. It provides further evidence that $\Sigma_{\mathrm{SFR}}$ is more important than total SFR on $\sigma_{\mathrm{gas}}$, as demonstrated in Figure \ref{fig:sigma_properties}.

\subsection{Non-rotational gas velocities}
\label{sec:non-rotational velocity of gas}

We also want to investigate other mechanisms that might influence gas turbulence, such as gas accretion and gas transportation through a disc. {\secRevise We use the velocity residual between the observed MAGPI velocity map, derived using single Gaussian component fit, and the convolved model} as an indicator of non-regular rotational velocity. The model velocity profile in \textsc{blobby3d} {\firstRevise only contains rotation}. Therefore, the deviation of the observed velocity map from the model velocity map can be considered as non-rotational {\secRevise motion} {\firstRevise (or at least deviations from regular rotation)}. We calculate the average velocity residual ($\Delta v$) weighted by H$\alpha$ line S/N of each spaxel {\teamRevise to reduce the impact of the low S/N spaxels at the edge of galaxy}, as follows: 
\begin{equation}
    \Delta v = \frac{\sum_{i\in \Omega }^{}\left |v_{\mathrm{con}}-v_{\mathrm{data}}  \right |_i\cdot sn_i}{\sum_{i\in \Omega }^{} sn_i}~,
\end{equation}
{\firstRevise where $\Omega$ includes all spaxels with H$\alpha$ S/N greater than 10, $v_{\mathrm{con}}$ is the velocity in {\secRevise the} convolved model of each spaxel, $v_{\mathrm{data}}$ is the velocity in {\secRevise the} data, $sn$ is the S/N of each spaxel. Then} we investigate its correlation with gas velocity dispersion. 

\begin{figure*}
    \centering
    \includegraphics [width=0.95\textwidth]{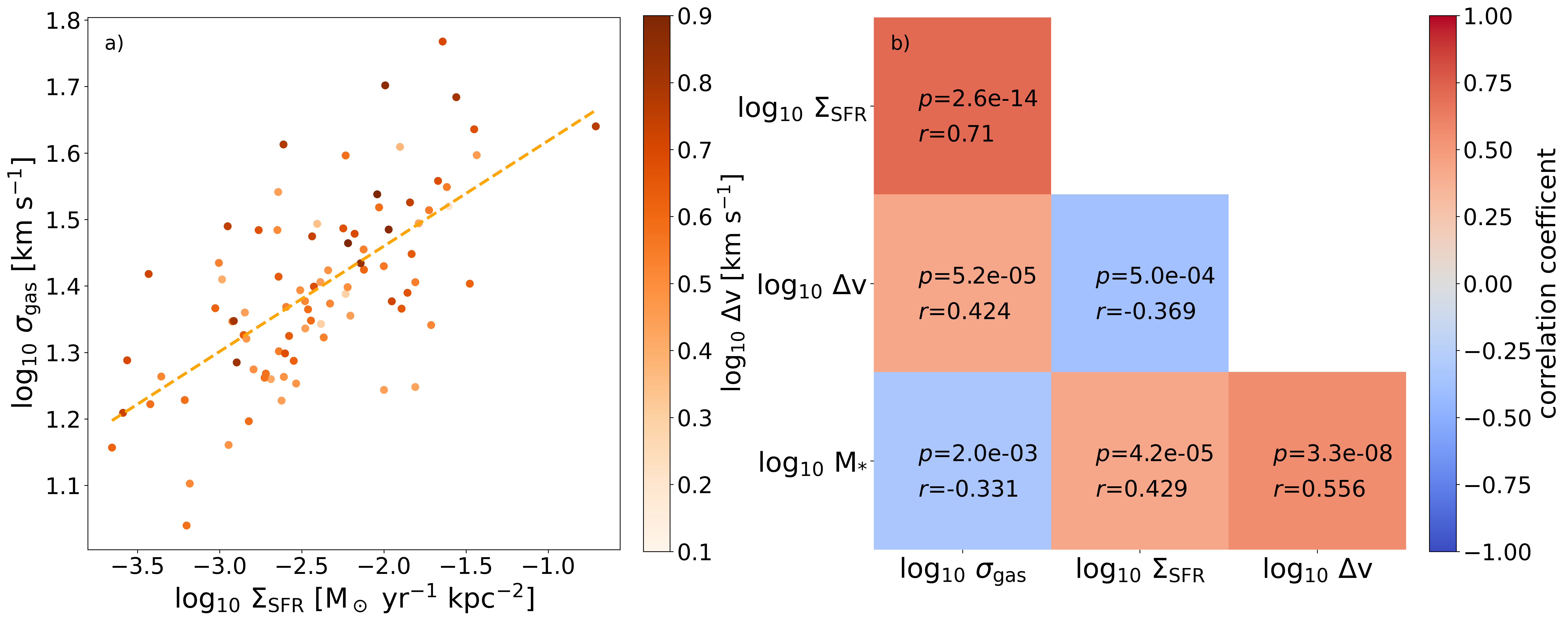}
    \caption{a) Gas velocity dispersion versus $\Sigma_{\mathrm{SFR}}$ colour-coded by average velocity residual ($\Delta v$). The dashed line is the ordinary least squares regression line. b) The Pearson partial correlation between $\sigma_{\mathrm{gas}}$, $\Sigma_{\mathrm{SFR}}$, $\Delta v$ and $M_{*}$. {\secRevise The Pearson correlation coefficient ($r$) and $p$-value are shown in each box, which is colour-coded by correlation coefficient.} The result shows that in addition to $\Sigma_{\mathrm{SFR}}$, gas velocity dispersion may also be related to non-{\secRevise circular} motions of the gas.}
    \label{fig:vel_res_combine}
\end{figure*}

Since the $\sigma_{\mathrm{gas}}-\Sigma_{\mathrm{SFR}}$ is the strongest correlation we find ($r=0.65$), we colour-coded the galaxies in our sample in the $\sigma_{\mathrm{gas}}-\Sigma_{\mathrm{SFR}}$ plane by their average velocity residual (shown in Figure \ref{fig:vel_res_combine}a), to see whether non-rotational velocity contributes to the scatter of this relation. {\secRevise The dashed line in Figure \ref{fig:vel_res_combine}a is fitted by ordinary least squares regression.} We find that the galaxies above the line tend to {\firstRevise have slightly (but non-negligible) higher} velocity residuals. The average velocity residual is $4.54 \pm 0.28$\,km\,s$^{-1}$ for the galaxies above the line and is $3.67 \pm 0.16$\,km\,s$^{-1}$ for the galaxies under the line. The velocity residual difference is small compared to the rotational velocity of the galaxy ($\sim 100$\,km\,s$^{-1}$). When we look at the large-scale velocity residual in individual galaxies, the velocity residual in some regions could be up to $\sim 40$\,km\,s$^{-1}$ (e.g. see Figure \ref{fig:m=3res} below).

We further investigate the partial correlation between $\sigma_\mathrm{gas}$, $\Sigma_{\mathrm{SFR}}$, $\Delta v$ and $M_{*}$, as shown in Figure \ref{fig:vel_res_combine}b. The correlation coefficient between $\sigma_{\mathrm{gas}}$ and $\Sigma_{\mathrm{SFR}}$ is $r = 0.71$. The correlation coefficient between $\sigma_{\mathrm{gas}}$ and $\Delta v$ is $r = 0.424$. The result shows that in addition to $\Sigma_{\mathrm{SFR}}$, {\secRevise our measurement of} gas velocity dispersion may also be related to non-{\secRevise circular} motions of the gas. \referee{We test an alternative method to find asymmetry, by rotating the seeing convolved velocity map by 180\textdegree, but this approach has more noise (the velocity residuals are consistently larger), likely because a non-uniform flux distribution plus beam smearing can generate kinematic asymmetries. This method reduces the significance of the signal we see, but we do not consider it reliable.}

Among all MAGPI galaxies in our sample, we find that four galaxies with high $\Delta v$ show an interesting Fourier $m=3$ pattern in the velocity residual maps. We will further discuss this in Section \ref{sec:drivers of gas turb}.

\subsection{Comparison to the SAMI and KROSS surveys}
\label{sec:compare sami and kross}

\begin{figure*}
    \centering
    \includegraphics [width=0.95\textwidth]{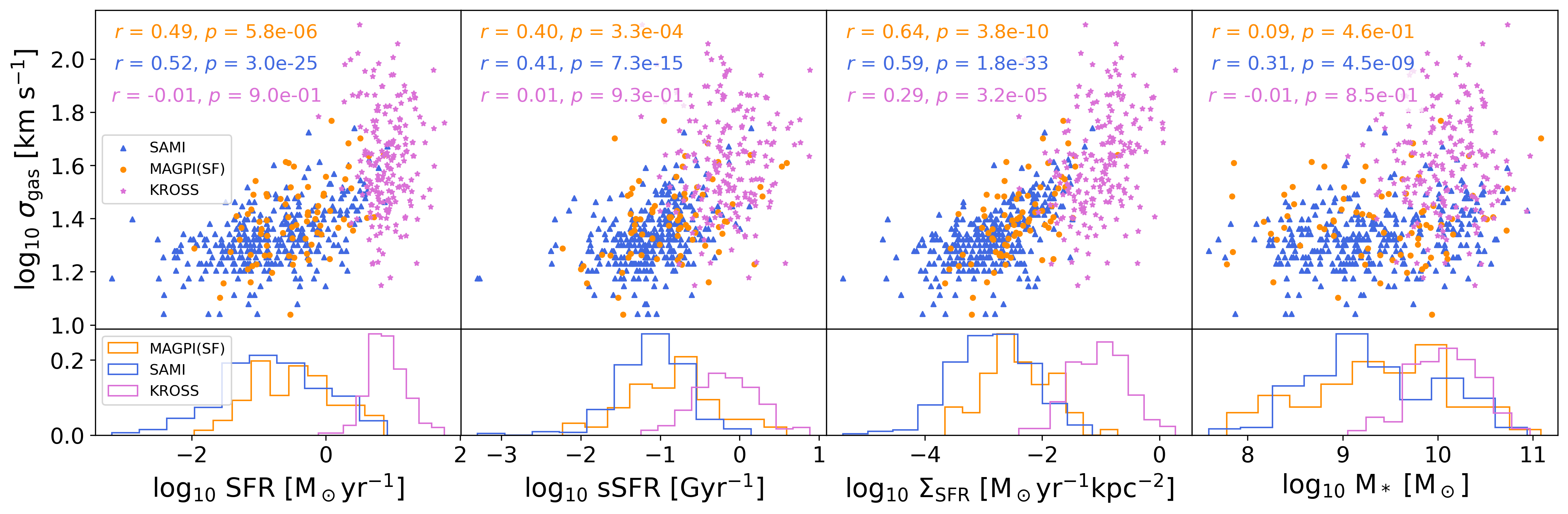}
    \includegraphics [width=0.95\textwidth]{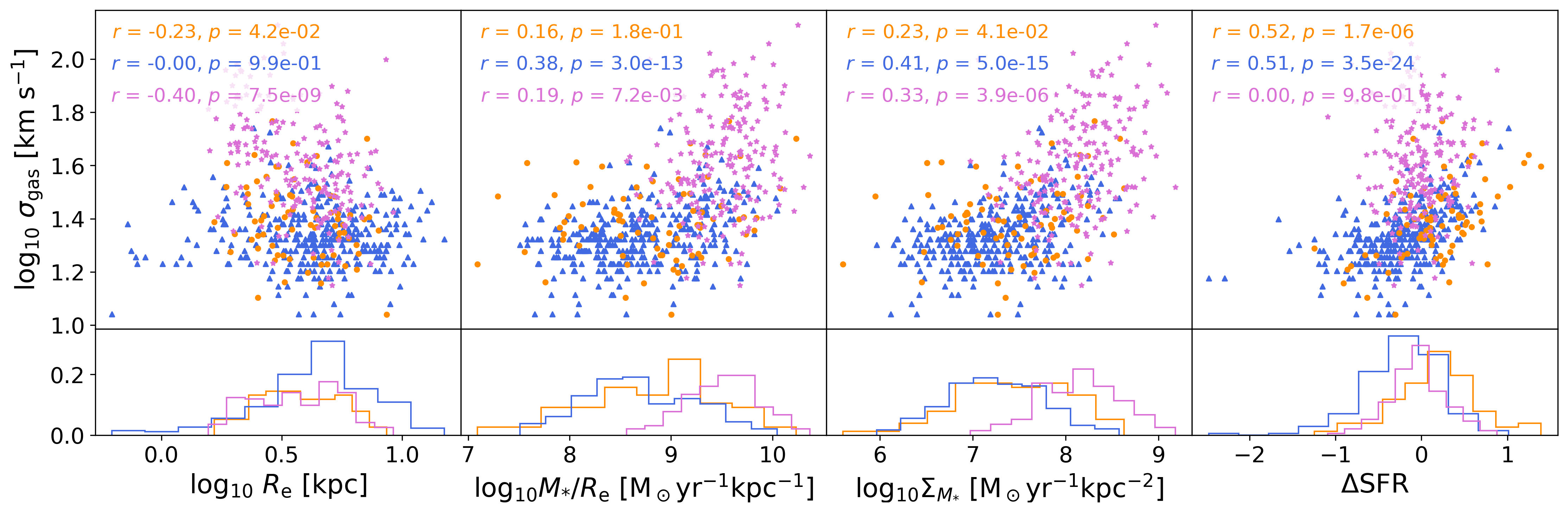}
    \caption{The relationship between gas velocity dispersion and galaxy properties for MAGPI (orange), SAMI (blue) and KROSS (purple) survey. The lower panel of each plot shows the distribution of galaxy properties. The Pearson correlation coefficients ($r$) and $p$-values are shown on each plot. Both MAGPI and SAMI galaxies have the strongest correlation between $\sigma_{\mathrm{gas}}$ and $\Sigma_{\mathrm{SFR}}$. KROSS galaxies also have a modest correlation between $\sigma_{\mathrm{gas}}$ and $\Sigma_{\mathrm{SFR}}$.}
    \label{fig:sigma-properties_3survey}
\end{figure*}

We compare the MAGPI sample with the SAMI ($z\sim 0.05$) and KROSS ($z\sim 1$) samples, of which the $\sigma_{\mathrm{gas}}$ is measured in a consistent method. As the SAMI and KROSS samples only include SF galaxies, we remove the AGN-host galaxies from our MAGPI sample in the following comparison. We note that removing AGN-host galaxies does not change our results. Figure \ref{fig:sigma-properties_3survey} compares the MAGPI, SAMI and KROSS velocity dispersions against several other galaxy properties. The Pearson correlation coefficients, $r$, and the $p$-values are shown for each survey separately. 
The histograms illustrate that the distribution of galaxy properties of MAGPI and SAMI galaxies are similar, although MAGPI galaxies have slightly higher stellar mass and higher SFR than SAMI galaxies. KROSS galaxies have $\sim 2$ dex higher SFR and $\sim 1$ dex higher stellar mass.

We find that both MAGPI and SAMI data show that $\sigma_\mathrm{gas}$ is strongly correlated with SFR and $\Sigma_{\mathrm{SFR}}$. Also, both MAGPI and SAMI data show that $\Sigma_{\mathrm{SFR}}$ has a higher correlation coefficient than SFR, which {\firstRevise may indicate} that higher energy injection per unit area by star-formation feedback, rather than total SFR of the galaxy, drives higher $\sigma_\mathrm{gas}$. {\firstRevise We also find $\Delta \mathrm{SFR}$ has a strong correlation ($r \sim 0.5$) with $\sigma_\mathrm{gas}$ for SAMI and MAGPI galaxies.}

Similar to Figure \ref{fig:partial_corr}, the partial correlations between $\sigma_{\mathrm{gas}}$, SFR, $M_*$ and $R_\mathrm{e}$ for SAMI galaxies show that there is no partial correlation between $\sigma_{\mathrm{gas}}$ and $M_*$, indicating that the relation between stellar mass and dispersion is driven by SFR for SAMI galaxies. There is no partial correlation between $\sigma_{\mathrm{gas}}$ and $M_*$ for KROSS galaxies. The progenitors of SAMI galaxies, at the redshifts of MAGPI and KROSS, will have lower stellar masses. However, the stellar mass has no partial correlation with $\sigma_{\mathrm{gas}}$ for SAMI, MAGPI and KROSS galaxies, indicating that mass-related progenitor bias is unlikely to be significant for our $\sigma_{\mathrm{gas}}$ analysis.

Interestingly, {\firstRevise although there is no correlation between $\sigma_\mathrm{gas}$ and SFR found for KROSS galaxies, there is a significant correlation between $\sigma_\mathrm{gas}$ and $\Sigma_{\mathrm{SFR}}$. This result suggests $\Sigma_{\mathrm{SFR}}$ is also more important than SFR for high redshift galaxies. However, we note that the range of SFR for KROSS galaxies {\secRevise is narrower (on a log-scale) than SAMI and MAGPI, which limits our ability to detect a correlation.}}

\begin{figure}
    \includegraphics [width=0.45\textwidth]{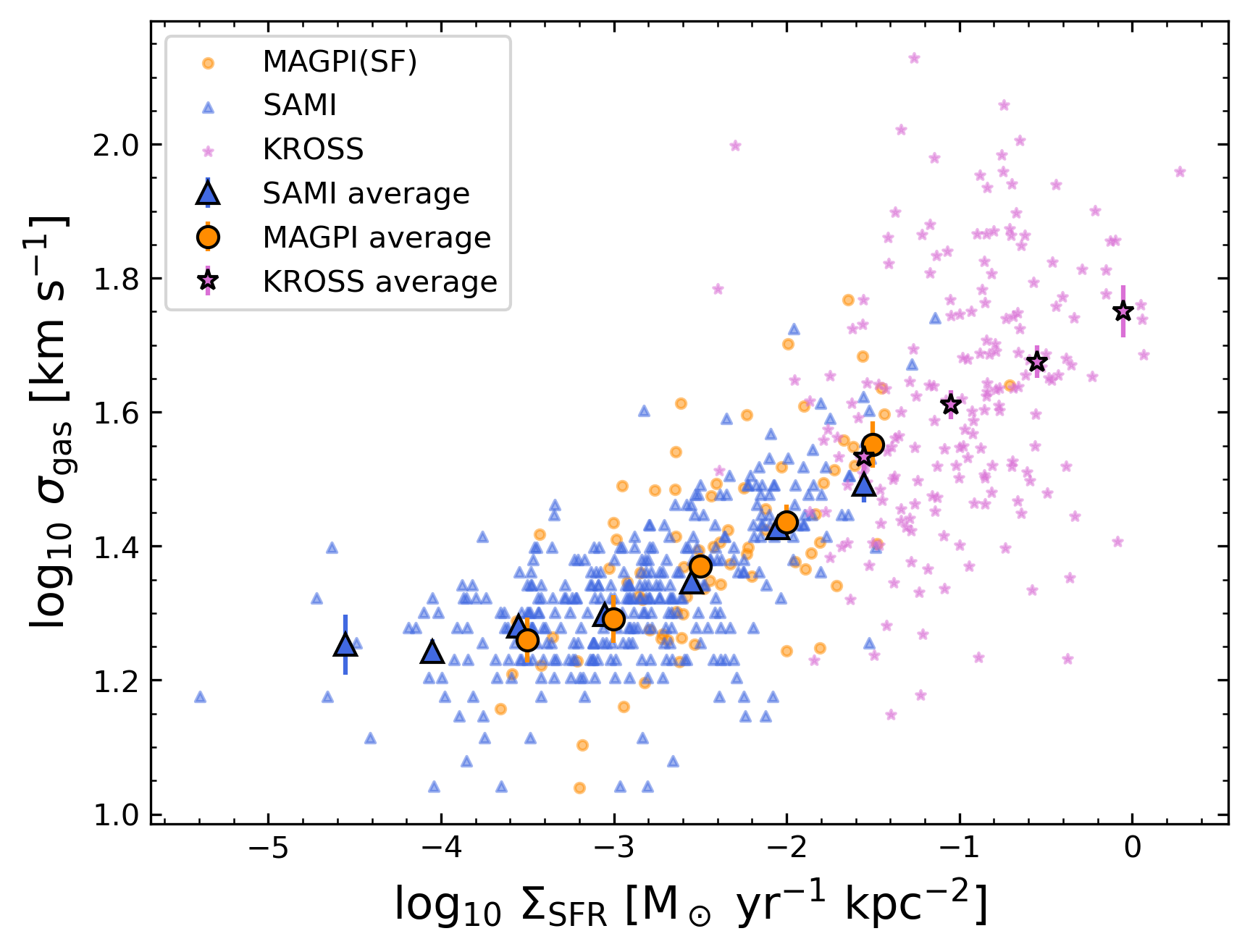}
    \caption{The $\sigma_{\mathrm{gas}}-\Sigma_{\mathrm {SFR}}$ relation for MAGPI (orange), SAMI (blue) and KROSS (purple) galaxies. The average $\sigma_{\mathrm{gas}}$ in each 0.5 dex bin for SAMI, MAGPI and KROSS are shown as black-edge triangles, circles and stars. The errorbars show the standard error of the mean.}
    \label{fig:sigma-Sigma_MAGPI_SAMI}
\end{figure}

As $\Sigma_{\mathrm{SFR}}$ has the strongest correlation with $\sigma_{\mathrm{gas}}$ for MAGPI and SAMI galaxies, and a moderate correlation with KROSS galaxies, we calculate the average $\sigma_{\mathrm{gas}}$ in each 0.5 dex bin for MAGPI, SAMI and KROSS galaxies separately. Figure \ref{fig:sigma-Sigma_MAGPI_SAMI} shows that the average $\sigma_{\mathrm{gas}}$ is similar for SAMI, MAGPI and KROSS galaxies at the fixed $\Sigma_{\mathrm{SFR}}$ and the average $\sigma_{\mathrm{gas}}$ lines up for galaxies from three surveys at different redshifts, which may suggest that the mechanism that relates to $\Sigma_{\mathrm{SFR}}$ is the main driver of gas turbulence at different redshifts. 

\section{DISCUSSION}
\label{sec:discussion}

\subsection{Gas velocity dispersion evolution through cosmic time}

\begin{figure*}
    \centering
    \includegraphics [width=0.8\textwidth]
    {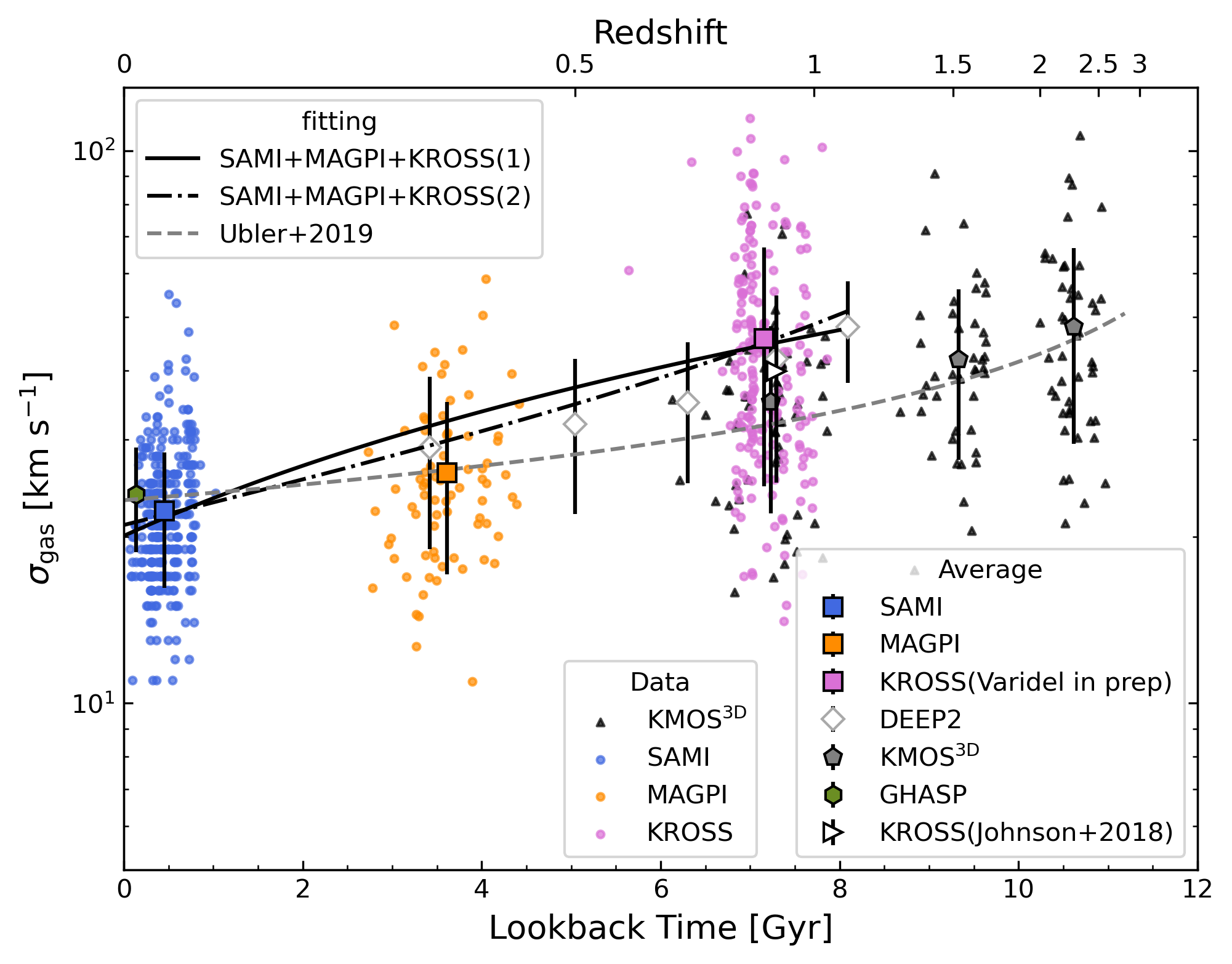}
    \caption{The gas velocity dispersion of SAMI, MAGPI and KROSS galaxies as a function of lookback time. The data from KMOS$^{\textup{3D}}$ \citep{Ubler:2019} are shown as black triangles. The average $\sigma_{\mathrm{gas}}$ and the standard deviation of SAMI \citep{Varidel:2020}, MAGPI, KROSS (Varidel et al. in prep), DEEP2 \citep{Kassin:2012}, KMOS$^{\textup{3D}}$ \citep{Ubler:2019}, GHASP \citep{Epinat:2010} and KROSS \citep{Johnson:2018} are shown as different black-edge markers. The black solid line shows the ordinary linear regression between gas velocity dispersion and lookback time for the combination of SAMI, MAGPI and KROSS data, {\secRevise for} which gas velocity dispersions are measured with the same method. The black dot-dash line shows the ordinary linear regression between gas velocity dispersion and redshift for the combined data. {\secRevise The grey dashed line is a robust $\sigma_\mathrm{gas}-z$ linear relation \citep{Cappellari:2013} from Figure. 6 in \citet{Ubler:2019}, fitted using the ionised gas velocity dispersion data from literature at $0<z<4$.} This plot shows that the gas velocity dispersions of SAMI, MAGPI and KROSS galaxies are consistent with the decreasing trend of gas velocity dispersion over cosmic time suggested by previous studies \citep{Kassin:2012,Wisnioski:2015,Ubler:2019}.}
    \label{fig:sigma-lookbacktime}
\end{figure*}

We show {\secRevise the} intrinsic gas velocity dispersion of SAMI, MAGPI and KROSS galaxies as a function of lookback time ($t_\mathrm{lookback}$) in Figure \ref{fig:sigma-lookbacktime}. The velocity dispersion of these galaxies is all measured using \textsc{blobby3d}. Previous observational studies have shown gas velocity dispersion decreases with cosmic time, so that galaxies have higher gas velocity dispersion when the Universe was younger \citep{Kassin:2012,Wisnioski:2015,Ubler:2019}. With the addition of the MAGPI data, a large IFS dataset at this epoch, we can explore the different mechanisms which may drive turbulence and their evolution. 

Two versions of the ordinary linear regression fitting of the gas velocity dispersions of SAMI, MAGPI and KROSS galaxies are shown in Figure \ref{fig:sigma-lookbacktime}. The linear relation between gas velocity dispersion and lookback time (line 1) is shown as a black solid line and is given by
\begin{equation}
    \sigma_\mathrm{gas}[\mathrm{km\:s^{-1}}] = 3.4(\pm 0.2)\cdot t_\mathrm{lookback}+20.1(\pm0.8)~.
\end{equation}
The relation between gas velocity dispersion and redshift (line 2) is shown as a black dot-dash line and is given by
\begin{equation}
    \sigma_\mathrm{gas}[\mathrm{km\:s^{-1}}] = 27.5(\pm 1.4)\cdot z+21.0(\pm 0.7)~.
\end{equation}
Our results show a decreasing trend of gas velocity dispersion from $z=1$ to $z=0$, which is consistent with previous studies. Both \citet{Wisnioski:2015} and \citet{Ubler:2019} showed clear decreasing trends of gas velocity dispersion over cosmic time using data of KMOS$^{\textup{3D}}$ in the redshift range $0.6<z<2.6$. \citet{Ubler:2019} did a linear regression of gas velocity dispersion and redshift {\secRevise using the gas velocity dispersion of KMOS$^{\textup{3D}}$ galaxies and ionised gas velocity dispersion from the literature at $0<z<4$}. {\secRevise The fit in \citet{Ubler:2019} is described by the equation}
\begin{equation}
    \sigma_\mathrm{gas}[\mathrm{km\:s^{-1}}] = 9.8(\pm 3.5)\cdot z + 23.3 (\pm 4.9)~.
\end{equation}

The gas velocity dispersion of SAMI, MAGPI and KROSS in our results are consistent with $\sigma_{\mathrm{gas}}-z$ linear fitting in \citet{Ubler:2019}. The gas velocity dispersion of MAGPI galaxies at $z\sim 0.3$ is in good agreement with the results in {\secRevise the slit based} DEEP2 survey \citep{Kassin:2012}. {\secRevise The average velocity dispersion of KROSS galaxies measured {\teamRevise with} \textsc{blobby3d} {\teamRevise agrees} with \citet{Johnson:2018}, who measured the gas velocity dispersion of KROSS galaxies {\teamRevise fitting a single Gaussian component with beam} smearing correction applied.} The KROSS galaxies at $z\sim 1$ have higher gas velocity dispersion than the KMOS$^{\textup{3D}}$ galaxies at a similar redshift. This difference may be {\teamRevise due to differences in sample selection}; \citet{Ubler:2019} applied strict criteria to select {\teamRevise only settled-disc galaxies}, {\teamRevise which may exclude the most turbulent galaxies and thus lower the average gas dispersion of the sample}. {\teamRevise We remark that we did not apply the strict selection criteria of \citet{Ubler:2019} to our sample of KROSS galaxies, but we removed galaxies with clear indications of ongoing major mergers.}

The evolution of gas velocity dispersion fits in a consistent picture with the evolution of star formation and gas fraction. Previous studies have found that the sSFR decreases over cosmic time \citep{Madau:1998,Hopkins:2006,Madau:2014,Leslie:2020,Popesso:2023}. The decreasing global averages of gas velocity dispersion is consistent with decreasing sSFR, which is an important galaxy property {\secRevise that} correlates with gas velocity dispersion. {\secRevise The evolution in sSFR is fairly flat from $z\sim 5$ to $z\sim 1$, only slightly decreasing. Most of the change appears to have happened between $z\sim 1$ to 0.} Galaxy gas content, which is the fuel of star formation, also evolves with cosmic time. Star-forming galaxies in earlier cosmic epochs held more molecular gas compared to the present time \citep[e.g.][]{Tacconi:2020}. The gas fractions of star-forming galaxies have decreased from $z=3$ to $z=0$ at a roughly constant rate \citep{Wiklind:2019}. 

\subsection{Drivers of gas turbulence}
\label{sec:drivers of gas turb}

We discussed the correlation between SFR and $\sigma_{\mathrm{gas}}$ ($r=0.46$) and between $\Sigma_{\mathrm{SFR}}$ and $\sigma_{\mathrm{gas}}$ ($r=0.65$) for MAGPI galaxies in Section \ref{sec:correlation_sigma_properties}. There is also a strong correlation between $\Sigma_{\mathrm{SFR}}$ and $\sigma_{\mathrm{gas}}$ for SAMI galaxies and a moderate correlation for KROSS galaxies. The average $\sigma_{\mathrm{gas}}$ is similar for SAMI, MAGPI and KROSS galaxies at the fixed $\Sigma_{\mathrm{SFR}}$, as shown in Figure \ref{fig:sigma-Sigma_MAGPI_SAMI}.

Star-formation feedback is considered to be an important source of gas turbulence since it injects energy into the interstellar medium \citep{Ostriker:2011,Shetty:2012,Faucher-Giguere:2013,Krumholz:2016,Krumholz:2018,Bacchini:2020,Sun:2020,Ceverino:2017}. Some analytical models and theories proposed that the energy from star-formation feedback sweeps up gas, then the shell of gas breaks up and merges with the interstellar medium, generating turbulence. The rate of energy injection per unit area, which is determined by the momentum injected per unit mass, star-formation surface density and gas velocity dispersion, balances the energy loss rate by turbulent dissipation \citep{Krumholz:2018,Faucher-Giguere:2013,Krumholz:2017}. {\secRevise Given} the importance of stellar feedback on driving turbulence, we expect a positive correlation between gas velocity dispersion and $\Sigma_{\mathrm{SFR}}$.

The strong correlations between $\sigma_{\mathrm{gas}}$ and SFR, and between $\sigma_{\mathrm{gas}}$ and $\Sigma_{\mathrm{SFR}}$ have been observed by many observational studies \citep{Green:2010,Lehnert:2013,Green:2014,Moiseev:2015,Zhou:2017,Johnson:2018,Yu:2019,Varidel:2020,Law:2022} and simulation studies \citep{Shetty:2012,Hung:2019,Jimenez:2022,Rathjen:2023}, that can be considered as the evidence of star-formation feedback driving gas turbulence. In particular, \citet{Egorov:2023} using high-resolution observations of gas superbubbles, showed a dependence of the kinetic energy of ionised gas velocity dispersion on energy input from stellar feedback, which provided further support {\secRevise for} the star-formation feedback energy injection theory.

{\secRevise The degree of star-formation feedback in increasing gas velocity dispersion} is controversial. Some studies proposed that star-formation feedback is the dominant factor driving gas turbulence. \citet{Faucher-Giguere:2013} and \citet{Bacchini:2020} built theoretical models to show that star-formation feedback alone is enough to sustain gas turbulence in local and high-redshift star-forming galaxies without other energy sources. On the contrary, some studies suggested a need for additional sources of gas turbulence \citep{Zhou:2017,Bik:2022,Forbes:2023}.

Another possible explanation for the strong correlation between $\sigma_{\textup{gas}}$ and $\Sigma_{\mathrm{SFR}}$ is {\secRevise that} gravitational instability drives gas velocity dispersion \citep{Krumholz:2016,Krumholz:2018,Molina:2020}. The Toomre-Q parameter \citep{Toomre:1964}, or gravitational stability parameter, can be approximated by
\begin{equation}
    Q\approx \sqrt{2}\frac{v_c \sigma _{\textup{gas}}f_\textup{g}}{\pi G r\Sigma _{\textup{gas}}}~,
\end{equation}
where $Q$ is the total Toomre-Q parameter for both gas and stars, $v_c$ is {\secRevise the galaxy rotation curve velocity (assuming a flat rotation curve)}, $f_\textup{g}$ is {\secRevise the} gas fraction, $\Sigma _{\textup{gas}}$ is {\secRevise the} gas surface density. The gravitational instability creates turbulence to keep the galaxy disc in marginally stable status, i.e. $Q\approx 1$. The energy source for this mechanism is gas inward transportation through the disc. If instability arises in the disc, it will break the axisymmetry, causing the subsequent torques that drive gas inward motion until marginal stability is restored. \citep{Bertin:1999,Krumholz:2010,Forbes:2012,Forbes:2014,Krumholz:2018}. The Kennicutt-Schmidt relation shows {\secRevise us} that $\Sigma_{\mathrm{SFR}}$ is proportional to $\Sigma _{\textup{gas}}$ \citep{Kennicutt:1998,Kennicutt:2012}. Therefore, the strong correlation we find between $\sigma_{\textup{gas}}$ and $\Sigma_{\mathrm{SFR}}$ may be a correlation between $\sigma_{\textup{gas}}$ and $\Sigma_{\mathrm{gas}}$ to keep the galaxy disc marginally stable, $Q\approx 1$. {\secRevise Further observations of $f_\mathrm{gas}$ and $\Sigma_{\mathrm{gas}}$ are needed to test this idea.} \referee{For example, \citet{Krumholz:2016} showed that the $\sigma_\mathrm{gas}$-SFR relation is different for gravity-driven and star formation feedback-driven models. In the gravity-driven model, $f_\textup{g}$ changes the slope of the $\sigma_\mathrm{gas}$-SFR relation.}

\begin{figure*}
    \centering
    \includegraphics [width=0.95\textwidth]{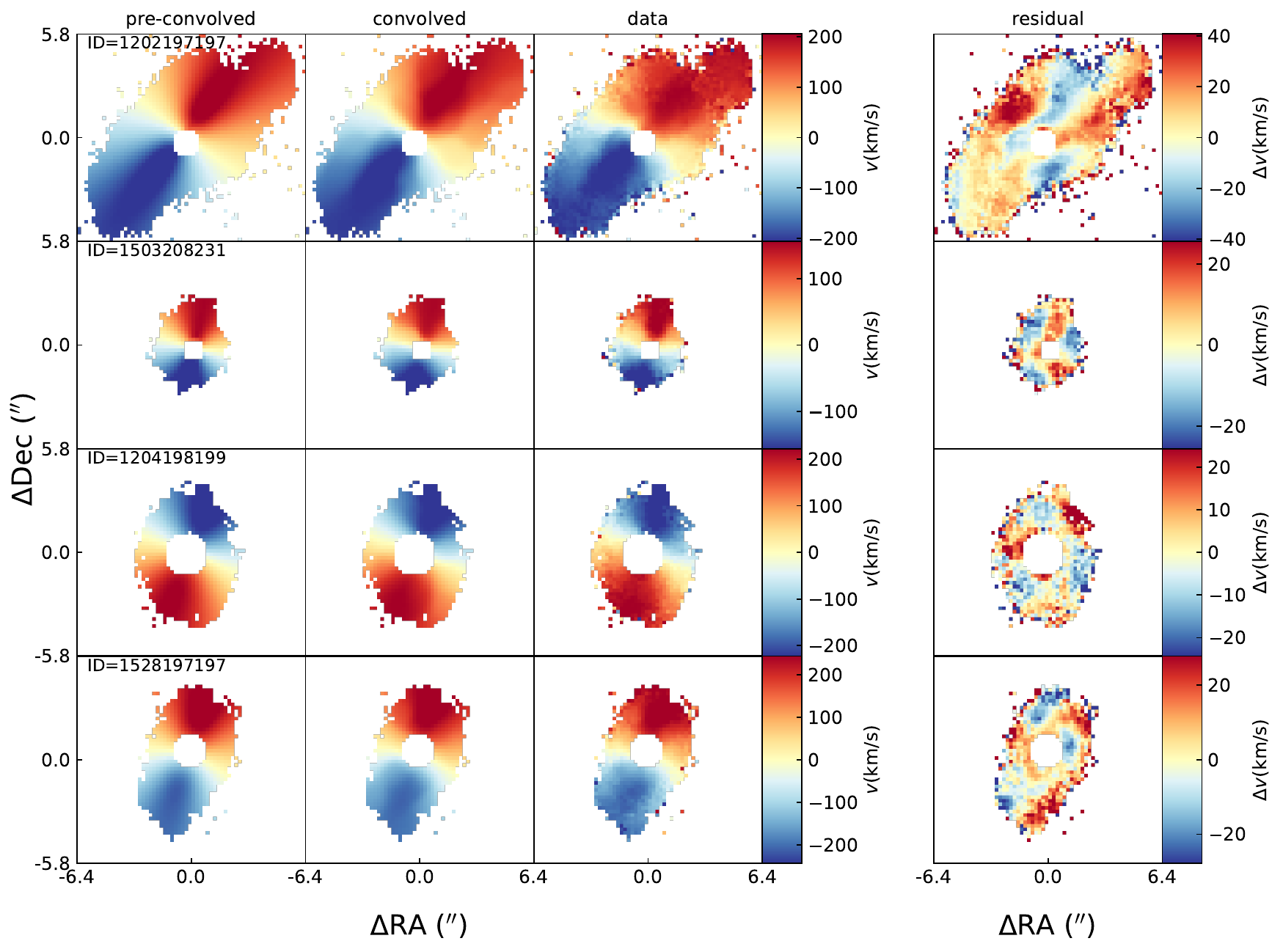}
    \caption{Four MAGPI galaxies that have $m=3$ Fourier mode in the velocity residual maps. The first column shows the model made by \textsc{blobby3d}; the second column shows the model convolved with the PSF and LSF of the data; the third column shows the results of the single Gaussian fit to the observed datacube; the fourth column shows the residuals between the data and the convolved model.}
    \label{fig:m=3res}
\end{figure*}

Gravitational instabilities can also be induced by external influences like galaxy-galaxy interactions or accretion. In Section \ref{sec:non-rotational velocity of gas}, we find some galaxies have large $\Delta v$ values, indicating non-circular motions. Four galaxies with high $\Delta v$ ($\Delta v=8.6\pm2.0$\,km\,s$^{-1}$) show an interesting Fourier $m=3$ pattern in their velocity residual maps, which means there are three approaching and three receding regions in velocity residual maps. Figure \ref{fig:m=3res} shows the four MAGPI galaxies with this pattern. These galaxies are AGN-host galaxies. We mask the AGN spaxels in these galaxies. We rule out the possibility of simple gas inflows for this pattern, as {\secRevise the simple gas inflow model} only creates {\secRevise an} axisymmetric pattern on {\secRevise the} line-of-sight velocity residual map \citep{Genzel:2023,Tsukui:2024}. The $m=3$ patterns we see may be caused by an interaction with a satellite galaxy, which disturbs the galaxy disc and triggers the bending wave or disc corrugation, which wraps up with differential rotation \citep[e.g.][]{Bland-Hawthorn:2021,TepperGarcia:2022}. A warp can be also caused by misaligned gas accretion and can provide kinetic energy to gas, increasing the velocity dispersion \citep{Khachaturyants:2022,Jimenez:2022}. One galaxy (ID=1204198199) shows evidence of interaction with neighbouring high mass galaxies in {\secRevise the} white light image {\secRevise from MUSE.} The other three galaxies do not have clear {\secRevise features indicating interactions} with other galaxies, but they live in groups that might have interaction with neighbouring low-mass galaxies. 

Another explanation for this $m=3$ pattern is spiral-arm related gas flows \citep{Kawata:2014,Grand:2016}. \citet{Grand:2016} showed that spiral arms induce large-scale streaming motion {\secRevise so} that the gas behind the arm moves azimuthally backwards and radially outwards, {\secRevise while} the gas at the front of the arm moves azimuthally forward and radially inward. The radial velocity {\secRevise in their model} is $\sim 20$\,km\,s$^{-1}$ and the azimuthal velocity is $\sim 10$\,km\,s$^{-1}$ relative to the mean rotational velocity. Although the line-of-sight velocity depends on the inclination of the galaxy and the location in the galaxy, the velocity residual we find is of the same order of magnitude as the combined effect of radial and azimuthal gas flows in \citet{Grand:2016}. Therefore, the $m=3$ pattern may {\secRevise be due to these galaxies having three spiral arms}. \citet{Bland-Hawthorn:2024} showed the gas-rich galaxies have a relatively shortlived feature of 3-arm ($m=3$) spirals using hydrodynamics simulations. They suggested that this feature is indicative of large-scale instability in the disc's early lifecycle. Two galaxies (ID=1202197197, 1204198199) with $m=3$ pattern are identified as three-armed spiral galaxies in \citet{Chen:2023}. {\secRevise Based on {\secRevise the white light MUSE image}, another one (ID=1528197197) {\secRevise has} some features of spiral arms, but {\secRevise the number of arms is uncertain. The remaining $m=3$ galaxy (ID=1503208231) does not} have a clear feature of arms.} Finally, for these $m=3$ featured galaxies, the possible influence of AGN-driven turbulence cannot be ruled out completely.

Mechanisms acting on different scales, such as stellar feedback, gas transportation and accretion, may contribute to the gas turbulence at the same time. However, simulating physical processes at different scales at the same time is difficult and simulations on different scales give different results. Using magnetohydrodynamic simulations of the star-forming interstellar medium in galaxy {\secRevise discs}, \citet{Rathjen:2023} showed the correlation between gas velocity dispersion and SFR vanished if they turned off stellar feedback. {\secRevise In contrast,} \citet{Jimenez:2022} showed that the correlation is weaker but still {\secRevise exists} in their EAGLE cosmological simulations without star-formation feedback. Note that these two studies investigate the relation in different scales. \citet{Rathjen:2023} focused on a galactic scale and did not include the impact of cosmological accretion on turbulence. On the contrary, \citet{Jimenez:2022} studied the impact of several mechanisms on turbulence on cosmological {\secRevise scales}, but their results were limited by {\secRevise the} spatial resolution of cosmological simulations {\secRevise and the} sub-grid physics {\secRevise that} was employed for stellar feedback.

The importance of different mechanisms may also vary with redshift and halo mass \citep{Ginzburg:2022,Jimenez:2022}. \citet{Ginzburg:2022} {\secRevise built an analytical model to show} that the contributions from three mechanisms to gas turbulence, including stellar feedback, gas transportation and gas accretion, evolved with cosmic time and were different for galaxies in different halo masses, but the relative contributions between them (the ratio of turbulent energy injection rates between different processes) were within a factor of $\sim 2$ in all discs at all redshifts. Their model predicts that stellar feedback is relatively more important for galaxies in lower halo mass and at lower redshift. The importance of gas transportation increases with redshift. Accretion is the most important mechanism for galaxies at $z>3$. Similarly, \citet{Kohandel:2020} used zoom-in simulations of galaxies to show accretion and mergers are the major sources of gas velocity dispersion for galaxies at $6<z<8$ and stellar feedback is sub-dominant.

Our results support a multi-drivers scenario. Both MAGPI and SAMI galaxies have a strong correlation between gas velocity dispersion and $\Sigma_{\mathrm{SFR}}$, agreeing with \citet{Ginzburg:2022}'s model that the gas velocity dispersion is predominantly supported by stellar feedback at low redshift. They also showed that the importance of transportation and accretion slowly increases with redshift, which is consistent with our result in Figure \ref{fig:vel_res_combine} that the gas velocity dispersion is correlated with the non-rotational motion of gas ($\Delta v$) for MAGPI galaxies and the correlation coefficient between $\Delta v$ and $\sigma_{\mathrm{gas}}$ ($r=0.424$) is smaller than that of $\Sigma_{\mathrm{SFR}}$ and $\sigma_{\mathrm{gas}}$ ($r=0.71$). \citet{Ginzburg:2022} also showed that for galaxies at $z\sim 1$, the contribution of gas transportation is similar to stellar feedback. Stellar feedback is relatively more important for galaxies in haloes with a halo mass log\,$M_{h,0}\lesssim 12.5$ at $z=0$, while gas transportation is more important for galaxies in haloes with log\,$M_{h,0}\gtrsim 12.5$. KROSS galaxies have $\sim 1$ dex higher stellar mass on average than MAGPI and SAMI galaxies, which implies {\secRevise that} they tend to live in higher mass {\secRevise haloes}. {\secRevise This fact may be why we see a weaker correlation} ($r=0.29$) between gas velocity dispersion and $\Sigma_{\mathrm{SFR}}$ for the KROSS galaxies and the scatter of gas velocity dispersion in the $\Sigma_{\mathrm{SFR}}$ bin is higher than SAMI and MAGPI. {\teamRevise However, we note that the range of $\Sigma_{\mathrm{SFR}}$ of KROSS galaxies on a log-scale is $\sim 1$ dex narrower than SAMI and MAGPI, which may limit our ability to detect the correlation. It is worth exploring the gas velocity dispersion of low mass high redshift galaxies.}

\section{CONCLUSIONS}
\label{sec:conclusion}

We studied the intrinsic H$\alpha$ kinematic properties of 110 galaxies at $z\sim0.3$ from the MAGPI survey using a {\secRevise 3D Bayesian inference} forward-modelling technique, \textsc{blobby3d}, which can model gas kinematics and gas substructure simultaneously. It assumes a regular rotating disc and decomposes {\secRevise the} ionised gas distribution into {\secRevise a number} of blobs with {\secRevise a} positive definite Gaussian basis function. The constructed 3D cubes are convolved by {\secRevise the} PSF and LSF to account for the beam smearing and instrumental broadening. The average gas velocity dispersion at $z\sim0.3$ is found to be $26.1\pm 8.7$\,km\,s$^{-1}$, where $\pm 8.7$\,km\,s$^{-1}$ represents the standard deviation.

We investigate the correlation between the gas velocity dispersion of MAGPI galaxies and galaxy properties. We find the gas velocity dispersion of MAGPI galaxies has the strongest correlation with $\Sigma_{\mathrm{SFR}}$ ($r=0.65$). The gas velocity dispersion also has a strong correlation with SFR ($r=0.46$) and with $\Delta \mathrm{SFR}$ ($r=0.50$). The partial correlation analysis between gas velocity dispersion, SFR, stellar mass and galaxy size shows that dispersion has a positive partial correlation with SFR and a negative partial correlation with size and {\teamRevise an insignificant} correlation with stellar mass. Using the residual between the regular rotating velocity map in the model and the velocity map of galaxies as an indicator of the non-rotational motion of the gas, we find the galaxies with higher velocity residual tend to have higher gas velocity dispersion.

We compare our results of MAGPI galaxies ($z\sim0.3$) with SAMI ($z\sim0.05$) and KROSS ($z\sim1$) survey, {\secRevise for} which the gas velocity dispersion is also measured using \textsc{blobby3d}. The $\sigma_{\mathrm{gas}}-$lookback time relation shows a decreasing trend of gas velocity dispersion over cosmic time. Both SAMI and MAGPI galaxies have a strong correlation between $\sigma_\mathrm{gas}$ and $\Sigma_\mathrm{SFR}$. KROSS galaxies have a moderate correlation between $\sigma_\mathrm{gas}$ and $\Sigma_\mathrm{SFR}$. We find that $\sigma_{\mathrm{gas}}$ is similar at fixed $\Sigma_\mathrm{SFR}$ for SAMI, MAGPI and KROSS galaxies, which suggests that the mechanisms related to $\Sigma_\mathrm{SFR}$ may be the dominant factor that drives gas turbulence from $z\sim1$ to $z\sim0$. The possible mechanisms that can explain this $\sigma_{\mathrm{gas}}$-$\Sigma_\mathrm{SFR}$ correlation are stellar feedback and gravitational instability.

When compared with theoretical models, our results support {\secRevise a} multi-drivers scenario. At $z\sim0.05$ and $z\sim0.3$, the star-formation feedback is the dominant driver of gas turbulence. The correlation between $\Delta v$ and $\sigma_{\mathrm{gas}}$ suggests that the non-rotational motion of the gas, such as gas transportation through the disc or gas accretion, also contributes to the gas turbulence for MAGPI galaxies at $z\sim0.3$. For KROSS galaxies at $z\sim1$, the importance of gas transportation may be higher than at low redshift, resulting in the moderate correlation between dispersion and $\Sigma_\mathrm{SFR}$ and higher scatter of $\sigma_{\mathrm{gas}}$ in the $\Sigma_\mathrm{SFR}$ bin.

\section*{Acknowledgements}

We wish to thank the ESO staff, and in particular the staff at Paranal Observatory, for carrying out the MAGPI observations. MAGPI targets were selected from GAMA. GAMA is a joint European-Australasian project based around a spectroscopic campaign using the Anglo-Australian Telescope. GAMA was funded by the STFC (UK), the ARC (Australia), the AAO, and the participating institutions. GAMA photometry is based on observations made with ESO Telescopes at the La Silla Paranal Observatory under programme ID 179.A-2004, ID 177.A-3016. The MAGPI team acknowledge support by the Australian Research Council Centre of Excellence for All Sky Astrophysics in 3 Dimensions (ASTRO 3D), through project number CE170100013. YM is supported by an Australian Government Research Training Program (RTP) Scholarship. CF is the recipient of an Australian Research Council Future Fellowship (project number FT210100168) funded by the Australian Government. CL, JTM and CF are the recipients of ARC Discovery Project DP210101945. KG is supported by the Australian Research Council through the Discovery Early Career Researcher Award (DECRA) Fellowship (project number DE220100766) funded by the Australian Government. Y.P. acknowledges the support from the National Science Foundation of China (NSFC) grant Nos. 12125301, 12192220, 12192222, and the science research grants from the China Manned Space Project with No. CMS-CSST-2021-A07. GS and KH acknowledge funding from the Australian Research Council (ARC) Discovery Project DP210101945. SMS acknowledges funding from the Australian Research Council (DE220100003). LMV acknowledges support by the German Academic Scholarship Foundation (Studienstiftung des deutschen Volkes) and the Marianne-Plehn-Program of the Elite Network of Bavaria. M.V. was supported by philanthropic funding from The Johnston Fellowship and from other donor(s) who are families affected by mental illness who wish to remain anonymous.

\section*{Data Availability}

The MUSE data used in this work are available on the ESO public archive. The reduced MAGPI datacubes and the emission line data products will be public in the MAGPI team data release (Mendel et al. in prep and Battisti et al. in prep).



\bibliographystyle{mnras}
\bibliography{drivers_of_gas_turbulence} 



\bsp	
\label{lastpage}
\end{document}